\documentclass{article}
\usepackage{cite}
\usepackage{graphicx} 
\usepackage{hyperref}
\usepackage{subcaption}
\usepackage[export]{adjustbox}
\usepackage[percent]{overpic}
\usepackage{graphbox}
\usepackage{longtable}
\usepackage{amsmath}
\usepackage{amsfonts}
\usepackage{amsthm}
\usepackage{amssymb}
\usepackage{gensymb}
\usepackage{mathrsfs}
\usepackage{xcolor}
\usepackage{pict2e}
\usepackage{multirow}
\usepackage{lineno}
\usepackage{authblk}

\newcommand{\dd}[1]{\mathrm{d}{#1}}

\title{Search for Steady and Flaring Neutrino Emission from Cosmic Sources Using the Complete ANTARES Dataset
}

\author[ ]{ANTARES Collaboration:} \author[1,2]{A.~Albert}
\author[3]{S.~Alves\thanks{Corresponding author}}
\author[4]{M.~Andr\'e}
\author[5]{M.~Ardid}
\author[5]{S.~Ardid}
\author[6]{J.-J.~Aubert}
\author[7]{J.~Aublin}
\author[7]{B.~Baret}
\author[8]{S.~Basa}
\author[7]{Y.~Becherini}
% ...

\affil[1]{\scriptsize{Universit\'e de Strasbourg, CNRS,  IPHC UMR 7178, F-67000 Strasbourg, France}}
\affil[2]{\scriptsize{Universit\'e de Haute Alsace, F-68100 Mulhouse, France}}
\affil[3]{\scriptsize{IFIC - Instituto de F\'isica Corpuscular (CSIC - Universitat de Val\`encia) c/ Catedr\'atico Jos\'e Beltr\'an, 2 E-46980 Paterna, Valencia, Spain}}
\affil[4]{\scriptsize{Technical University of Catalonia, Laboratory of Applied Bioacoustics, Rambla Exposici\'o, 08800 Vilanova i la Geltr\'u, Barcelona, Spain}}
\affil[5]{\scriptsize{Institut d'Investigaci\'o per a la Gesti\'o Integrada de les Zones Costaneres (IGIC) - Universitat Polit\`ecnica de Val\`encia. C/  Paranimf 1, 46730 Gandia, Spain}}
\affil[6]{\scriptsize{Aix Marseille Univ, CNRS/IN2P3, CPPM, Marseille, France}}
\affil[7]{\scriptsize{Universit\'e Paris Cit\'e, CNRS, Astroparticule et Cosmologie, F-75013 Paris, France}}
\affil[8]{\scriptsize{Aix Marseille Univ, CNRS, CNES, LAM, Marseille, France }}
% ...

\begin{document}

\date{}
\maketitle

\begin{abstract}

ANTARES, a neutrino detector located in the depths of the Mediterranean Sea, operated successfully for over 15 years before being decommissioned in 2022. The telescope offered an ideal vantage view of the Southern Sky and benefited from optimal water properties for enhanced angular resolution. 
This study makes use of data collected over the entire operational period of  ANTARES to search for sources of high-energy cosmic neutrinos, considering both steady and flaring emission scenarios.
First, a time-integrated search for high-energy neutrino clustering across the celestial sphere is conducted. The most significant accumulation is found at coordinates $(\alpha, \delta) =(200.5^\circ\!, 17.7^\circ)$ with a post-trial p-value equal to 0.38. A dedicated search in the Galactic Plane is also performed for extended sources, yielding no significant excess. Additionally, a list of potential neutrino sources are investigated. The blazar \hbox{MG3 J225517+2409} is identified as the most significant object, yet the excess remains compatible with background fluctuations. A mild local excess of 2.4$\sigma$ is found for the blazar \hbox{TXS 0506+056}.
The full sky is also examined for the presence of flaring neutrino emissions. The most significant excess in this case corresponds to a $\sim$4-day flare from the direction $(\alpha, \delta) = (141.3^\circ\!, 9.8^\circ)$, with a post-trial p-value of 0.30.
Finally, the directions of sources highlighted in IceCube’s time-dependent searches are investigated. Temporal overlaps between ANTARES and IceCube flares are identified for \hbox{PKS 1502+106} and \hbox{TXS 0506+056}, with an estimated chance probability of about 0.02\%, making this observation particularly noteworthy.  

\end{abstract}

\section{Introduction}

Neutrinos, the elusive messengers of the cosmos, offer a unique window into the most extreme environments in the Universe and provide a valuable opportunity to solve the long-standing mystery of the origin of cosmic rays. After proving the existence of a diffuse neutrino flux~\cite{IceCube:2020wum, Abbasi:2021qfz, IceCube:2020acn}, in recent years, the IceCube Collaboration has made notable progress in the study of neutrinos: from pinpointing sources like TXS~0506+056~\cite{IceCube:2018cha, IceCube:2018dnn} and NGC~1068~\cite{IceCube:2022der}, to mapping the neutrino emissions from the Galactic Plane~\cite{IceCube:2023ame}. These observations have significantly reshaped our understanding of high-energy neutrino astrophysics and have motivated further investigations by complementary observatories worldwide.

The ANTARES neutrino telescope, situated in the Mediterranean Sea, offered a valuable perspective of the high-energy neutrino Universe from the Northern Hemisphere, with enhanced visibility of the Southern Sky, including most of the Galactic Plane. In fact, the ANTARES data have also revealed a mild excess along the Galactic Ridge~\cite{ANTARES:2022izu}, consistent with the IceCube findings. Moreover, although ANTARES did not collect enough events to resolve the diffuse astrophysical neutrino flux, its good sensitivity in the TeV range allowed to set a constraint for the unbroken power-law hypothesis, supporting IceCube's preference for a spectral break around the 10~TeV to 30~TeV range~\cite{ANTARES:2024ihw}. Additionally, the good optical properties of the water at the ANTARES site provided the detector with an excellent angular resolution, which enhanced its capability to detect and study neutrino sources. 

In this paper, the results of several searches for cosmic sources of neutrinos using a dataset spanning the full ANTARES data-taking period are presented. The paper is structured as follows: Section~\ref{sec:telesocpe} provides a description of the detector and of the data sample used in this work, while the analysis methods employed are described in Section~\ref{sec:analysis}. Section~\ref{sec:TimeIntResults} provides the results of the time-integrated studies, specifically, of a full-sky scan (~\ref{subsec:time-int-full-sky}), of a search for extended sources located along the Galactic Plane (~\ref{subsec:gal-plane}), and of a survey of a pre-selected list of astrophysical objects (~\ref{subsec:candidate-list}). The results of the time-dependent studies are presented in Section~\ref{sec:TimeDepResults}. Finally, conclusions are given in Section~\ref{sec:concl}.

\section{The ANTARES neutrino telescope and data sample} \label{sec:telesocpe}

ANTARES was an undersea high-energy neutrino detector~\cite{antaresdetector} situated 40 kilometres off the coast of Toulon, France, submerged beneath the Mediterranean Sea. After running for 15 years in its full configuration, it concluded its observation duty in February 2022. The telescope comprised a three-dimensional array of 885 photomultiplier tubes (PMTs) distributed along 12 vertical lines, each extending 450 meters in length, instrumenting a total volume of approximately $0.016\ \textrm{km}^3$. The lines were securely anchored to the seabed at a depth of around $2500\ \textrm{m}$ and kept taut by buoys at the top. Within pressure-resistant optical modules~\cite{ANTARES:2001nhp}, the PMTs collected Cherenkov photons induced in the medium by relativistic charged particles resulting from neutrino interactions occurring inside or nearby the instrumented region. The data obtained from the position, time, and charge collected in the PMTs constitute \textit{hits}~\cite{ANTARES:2006gto}, which were then used to deduce the direction and energy of the incident neutrino.

Two primary event topologies attributed to different neutrino flavours and interaction types could be distinguished: track-like and shower-like events~\cite{ANTARES:2020bhr}. Muon-neutrino charged current (CC) interactions produce relativistic muons capable of traversing extensive distances through the medium, thereby producing a track-like signature in the detector~\cite{ANTARES:2011vtx, ANTARES:2013tra}. The direction of the originating neutrino for well-reconstructed tracks was determined with a median angular resolution of $\sim$$0.8^{\circ}$ at neutrino energies around 1 TeV and of $\sim$$0.4^{\circ}$ for energies exceeding 10~TeV~\cite{ANTARES:2017dda}, thanks to the long lever arm of this channel. On the other hand, shower-like events arise from all-flavour neutral current (NC) interactions and ${\nu_e}$ and ${\nu_\tau}$ CC interactions\cite{ANTARES:2017ivh}. Characterised by nearly spherical light emission around the interaction point, with a few meters of elongation, this topology leads to a less accurate estimation of the parent neutrino direction when compared to the track channel. A median angular resolution of $\sim$$3^{\circ}$ is achieved for well-reconstructed showers with energies ranging from 1~TeV to 0.5~PeV~\cite{ANTARES:2017dda}. A wide overview of the ANTARES detector, operations, and scientific results is presented in~\cite{ANTARES:2025exu}.

Events recorded during the full data-taking period, i.e.\,from January 29, 2007, to February 13, 2022,  totalling 4541 days of livetime, are employed in this analysis. They are selected using the criteria outlined in~\cite{ANTARES:2017dda}, tailored to minimise the required neutrino flux for a $5\sigma$ detection of a point-like source exhibiting a $\propto E^{-2.0}$ emission spectrum. The selection includes cuts on the zenith angle, on the angular error estimate, and on parameters reflecting the reconstruction quality. In the shower channel, an additional criterion requires the interaction vertex to lie within a fiducial volume slightly exceeding the instrumented volume. Recent enhancements in calibrations have been employed to reconstruct all selected ANTARES events, resulting in slight variations in the reconstructed direction -- approximately 98\% of the selected events maintain a reconstructed direction within $1^{\circ}$ of their previous estimation -- and in the quality parameters. Furthermore, a refined version of the energy estimator for track-like events has been introduced to account for the time evolution of the detector~\cite{ANTARES:K40} along the entire data acquisition period. Shower-like events also benefit in this work from an improved energy estimation based on reconstruction parameters~\cite{ANTARES:2017ivh}.
A total of 11029 track-like events and 200 shower-like events are employed in this work. The estimated contamination from atmospheric muons is below 10\% for the track channel and below 18\% for the shower channel.

\section{Analysis method} \label{sec:analysis}

In this study, a model describing neutrino emissions following a simple power law is considered:
\begin{equation}\label{eq:FluxParam}
\frac{\dd{\phi_\nu}}{\dd{E}} = \Phi^{\nu+\bar{\nu}}_{1\rm\, GeV} \times \left( E_\nu \,/\, {1\rm \,GeV} \right)^{-\gamma}  \quad {\rm [GeV^{-1}cm^{-2}s^{-1}]} \,,
\end{equation}
where $\Phi^{\nu+\bar{\nu}}_{1\rm\, GeV}$ is the one-flavour neutrino flux normalisation factor at Earth (flavour equipartition is assumed). The spectral features of the signal flux depend solely on the spectral index $\gamma$, which is expected to be significantly harder than that of the atmospheric background ($\gamma_{\rm atm} \simeq 3.7$). As a result, the energy distribution of detected events allows to discriminate between signal and background.
Furthermore, neutrinos originating from a single astrophysical source are expected to cluster spatially around the source direction, in contrast to the isotropic distribution of atmospheric background events. When the signal emission is also time-dependent, the arrival time of the events becomes an additional tool for suppressing background contamination.

In order to exploit these distinguishing features, an unbinned likelihood function is built:
\begin{equation}\label{eq:likelihood}
   \log \mathcal{L} = \sum_{j}^{N_{\text{sam}}} \sum_{i}^{N_j} \log \left[ \frac{\mu^j_{\text{sig}}}{N_j} \mathcal{S}^j_i + (1 - \frac{\mu^j_{\text{sig}}}{N_j})\mathcal{B}^j_i \right] \,,
\end{equation}
where $j$ represents the samples considered in the analysis, i.e.\,tracks and showers, $N_{\text{sam}}$ the total number of samples, $i$ the events within the sample, $N_j$ the total number of events observed in the $j$-th sample,  and $\mu^j_{\text{sig}}$ the number of signal events in the $j$-th sample. 
The terms $\mathcal{S}_i^j$ and $\mathcal{B}_i^j$ are the probability density functions (PDFs) that describe the probability of an event being signal or background, respectively. Their detailed definitions are provided below.
The likelihood is maximised for the total number of signal events, $\mu_{\text{sig}}$, related to $\mu^j_{\text{sig}}$ as $\mu^j_{\text{sig}} = \mu_{\text{sig}} C^j$, with $C^j$ being the expected relative contribution of the given sample to the total detected cosmic neutrino flux. $C^j$ is derived from Monte Carlo simulations of astrophysical neutrinos for the assumed spectral index and is a function of the source declination, as detailed in~\cite{ANTARES:2020srt}. 

\subsection{Time-integrated approach}
\label{subsec:method-time-int}

In this case, the signal and background PDFs are built as the product of a directional and an energy term. 

The signal spatial PDF is obtained from a smooth parametrisation of the point spread function (PSF), derived from Monte Carlo simulations of cosmic neutrinos assuming a power-law energy spectrum. 
The PSF is derived as the probability density of the angular distance between the simulated and the reconstructed neutrino direction, $\Psi$, per unit solid angle $\Omega$: 
\begin{equation}\label{eq:PSF}    
\textrm{PSF}(\Psi) = \frac{\dd{\rm P}(\Psi)}{\dd{\Omega}} = \frac{\dd{\Psi}}{\dd{\Omega}}\frac{\dd{\rm P}(\Psi)}{\dd{\Psi}}= \frac{1}{2\pi \sin\Psi} \frac{\dd{\rm P}(\Psi)}{\dd{\Psi}} \,.
\end{equation}
In other words, for each ANTARES $i$-th event, $\textrm{PSF}(\Psi_i)$ provides the probability density of reconstructing the event within a radius $\Psi_i$ around the source. Under the assumption of the background being uniform in right ascension, the probability density function of finding a background event at a given direction of the sky is 
\begin{equation}\label{eq:sindec}
P^{\text{Space}}_\text{Bg}(\sin\delta_i) = \frac{1}{2\pi}R(\sin\delta_i) \,,
\end{equation}
with $R(\sin\delta)$ being the observed distribution of the sine of the declination divided by the total number of events in the data sample. Since the sample is dominated by atmospheric events, this PDF is directly built from the data.
Monte Carlo simulations of power-law energy spectrum cosmic neutrinos and of atmospheric neutrinos using the spectrum described in~\cite{Honda:2006qj} are used to derive the signal and background energy PDFs, $P^{\text{Energy}}_\text{Sg}(E_i)$ and $P^{\text{Energy}}_\text{Bg}(E_i)$, respectively. 

Therefore, the signal and background terms for the likelihood function shown in Eq.~\eqref{eq:likelihood} are built for tracks as:
\begin{align}
   \label{Eq:Sg_tracks} \mathcal{S}^{\rm Tracks}_i &= \text{PSF}(\Psi_i\, | \, \gamma) \times P^{\text{Energy}}_\text{Sg}(E_i\,|\,\delta_i, \gamma)\,, \\
    \label{Eq:Bg_tracks} \mathcal{B}^{\rm Tracks}_i &= P^{\text{Space}}_\text{Bg}(\sin\delta_i) \times P^{\text{Energy}}_\text{Bg}(E_i\,|\,\delta_i)\,,
\end{align}
where the dependency of the energy estimator, $E_i$, on the event declination, $\delta$, is considered in both the signal and background cases.  Shower events, which suffer from too low statistics to use declination dependent distributions, are evaluated with declination independent PDFs: 
\begin{align}
    \label{Eq:Sg_showers}\mathcal{S}^{\rm Showers}_i &= \text{PSF}(\Psi_i\, | \, \gamma) \times P^{\text{Energy}}_\text{Sg}(E_i\,|\, \gamma)\,, \\
    \label{Eq:Bg_showers}\mathcal{B}^{\rm Showers}_i &= P^{\text{Space}}_\text{Bg}(\sin\delta_i) \times P^{\text{Energy}}_\text{Bg}(E_i)\,.
\end{align}

Regarding the signal spectral index, two hypotheses are tested: $\gamma = 2.0$, predicted by the Fermi acceleration mechanism, and $\gamma = 2.5$, motivated by the recent IceCube neutrino observatory measurements~\cite{IceCube:2020acn,IceCube:2020wum,IceCube:2024fxo}.

Besides, if the hypothesis to be tested considers that the target source has a certain extension (the emission is not point-like), the PSF is modified to include this information. 
The emission profile of the sources is parametrised as a symmetrical 2D Gaussian function:

\begin{align}
    \label{Eq:2DGuassian}
    \mathcal{G}(\Delta\Upsilon,\sigma_\text{Ext}) = \frac{1}{2\pi\sigma_\text{Ext}^2}\times \exp{\big( -\frac{\Delta\Upsilon^2}{2\sigma_\text{Ext}^2} \big)}\,,
\end{align}
with $\Delta\Upsilon$ being the angular distance from the source centre and $\sigma_\text{Ext}$ being the source extension as provided by electromagnetic data.  The PDF $\dd{\rm P}(\Psi)/\dd{\Psi}$ is convolved with $\mathcal{G}(\Delta\Upsilon,\sigma_\text{Ext})$, and the resulting distribution is then used to obtain the appropriate PSF as shown in Eq.~\eqref{eq:PSF}.

At each investigated sky location, the likelihood is maximised with respect to the total number of signal events
and a test statistic is computed. The test statistic is defined as follows:
\begin{equation}\label{eq:TS}
    \mathcal{Q} = - 2 \times \log\left(\frac{\mathcal{L}(\mu_{\rm sig} = 0)}{\mathcal{L}(\mu_{\rm sig} = \hat{\mu}_\mathrm{sig})} \right)\,,
\end{equation}
that is, the difference between the value of the logarithm of the likelihood at the best-fit free parameter, $\hat{\mu}_\mathrm{sig}$, and the one for the background-only hypothesis. Distributions of the test statistic are built from pseudo-experiments (PEs), i.e.\, evaluating the likelihood on mock data samples composed only of background-like events which maintain the properties of the declination, right ascension, and energy distributions of real data. The fraction of background-only PEs with a value of $\mathcal{Q}$ larger than the one obtained with the data gives the significance (p-value) of each investigated sky location. In the case of a null or insignificant observation, 
upper limits on the flux normalisation $\Phi^{\nu+\bar{\nu}}_{1\rm\, GeV}$ are set.

When multiple sky locations are tested, the p-value observed at each direction is referred to as 
the \textit{pre-trial} p-value, which must be corrected to account for the ``look-elsewhere effect''~\cite{Lyons_2008}. To do so, the same analysis is repeated over many sky maps composed of only background-like events. For each PE, the lowest p-value obtained among all 
investigated locations is recorded, yielding a distribution of minimum p-values. Using this distribution, the probability of obtaining from background fluctuations a lower p-value than the one observed in the data is computed, which corresponds to the post-trial p-value.

Regarding systematic uncertainties, this work incorporates the same sources of uncertainty as those considered in~\cite{ANTARES:2017dda}, namely those related to: angular resolution, absolute pointing accuracy, background rate, and acceptance.

\subsection{Time-dependent approach}
\label{subsec:method-time-dep}

In this case, the tested hypothesis is that neutrinos arrive at the detector following a flaring episode in which the central time of flare and its duration are, \textit{a priori}, unknown. The signal and background terms of the likelihood -- Eq.~\eqref{Eq:Sg_tracks} and~\eqref{Eq:Bg_tracks} for tracks and Eq.~\eqref{Eq:Sg_showers} and~\eqref{Eq:Bg_showers} for showers -- are multiplied by a time-dependent term, following the strategy used in~\cite{ANTARES:2023lck}.
The background time PDF is the rate of background events that are recorded by the detector. It is built from data using a less stringent selection so that the resulting distribution is not dominated  by statistical fluctuations.  In order to prevent any exceptional deviation between the time distribution of the final sample and that of the looser selection from mimicking a signal, data-taking runs exhibiting such deviations are excluded, resulting in a slightly reduced track sample of 11023 events.
The signal time PDF, which is identical for both track and showers, is characterised by two generic time profiles, with a Gaussian-like and a box-like shape:
\begin{align}
    &P_\text{Sg}^\text{Gauss}(t_i) = \frac{1}{\sqrt{2\pi}\sigma_t}\times \exp{\big( -\frac{(t_i - T_0)^2}{2\sigma_t^2} \big)}\,,
    \label{eq:time_gauss} \\
    &P_\text{Sg}^\text{Box}(t_i) = \begin{cases}
    \frac{1}{2\sigma_{t}}\,, & \text{if $[T_{0} - \sigma_{t}] \leq t_i \leq [T_{0} + \sigma_{t}]$} \,;\\
    0\,, & \text{otherwise} \,; \label{eq:time_box}
  \end{cases}
\end{align}
with $t_i$ being the detection time of the event, and $T_0$ and $\sigma_t$ being the central time of the neutrino emission and the duration of the neutrino flare, respectively. 
Both $T_0$ and $\sigma_t$ are free parameters in the likelihood maximisation, with $\sigma_t$ taking values between $1$~day and $1000$~days, while $T_0$ spans over the time range of the investigated ANTARES data. 
Additionally, the spectral index $\gamma$, entering the likelihood through Eq.~\eqref{Eq:Sg_tracks} and Eq.~\eqref{Eq:Sg_showers}, is also left as a free parameter that can take values between 1.5 and 3.0.

The likelihood $\mathcal{L}(\mu_\text{sig}, \gamma, T_0, \sigma_t)$ is then maximised for each targeted sky direction. To prevent the bias of the fit for short duration flares, which are easier to accommodate in a given time range than longer ones, the test statistic is corrected as explained in~\cite{braun2010time}:
\begin{equation} \label{eq:TimeTS}
    \mathcal{Q} = - 2 \log \left( \frac{\Delta T}{\hat{\sigma}_t} \times \frac{\mathcal{L}(\mu_{\rm sig} = 0)}{\mathcal{L}(\mu_{\rm sig} = \hat{\mu}_\mathrm{sig})} \right),
\end{equation}
where $\Delta T$ is the allowed time range for $T_0$, while $\hat{\sigma}_t$ is the best-fit value of the flare duration.

Finally, pre-trial and post-trial p-values are computed using the same PE-based procedure as employed in the time-integrated analysis.

\section{Results of the time-integrated studies} \label{sec:TimeIntResults}

\subsection{Full-sky search} \label{subsec:time-int-full-sky}

In this search, the entire celestial sky beneath 40$^\circ$ in declination, i.e.\, the ANTARES visibility threshold when using upgoing events, is subdivided into a grid with a finer spacing than the typical event angular uncertainty.   The grid consists of a total of 2,583,552 pixels covering equal areas in solid angle and measuring approximately $(0.11^\circ  \times  0.11^\circ)$, obtained using the HEALPix pixelisation\footnote{\url{http://healpix.sourceforge.net}} tool with $N_{\textrm{side}}$~=~512 ~\cite{Zonca:2019vzt, Gorski:2004by}.  The likelihood function of Eq.~\ref{eq:likelihood} is then recursively evaluated assuming that each studied direction, determined by the coordinates of the centre of the pixel, is a potential location of a point-like source with $\gamma = 2.0$. Since this approach assumes that a neutrino source could be located anywhere in the sky, independently of any associated electromagnetic counterpart, this method ensures that the search remains unbiased by electromagnetic observations.

\begin{figure*}[h!]
\centering
    \begin{overpic}[width=\textwidth]{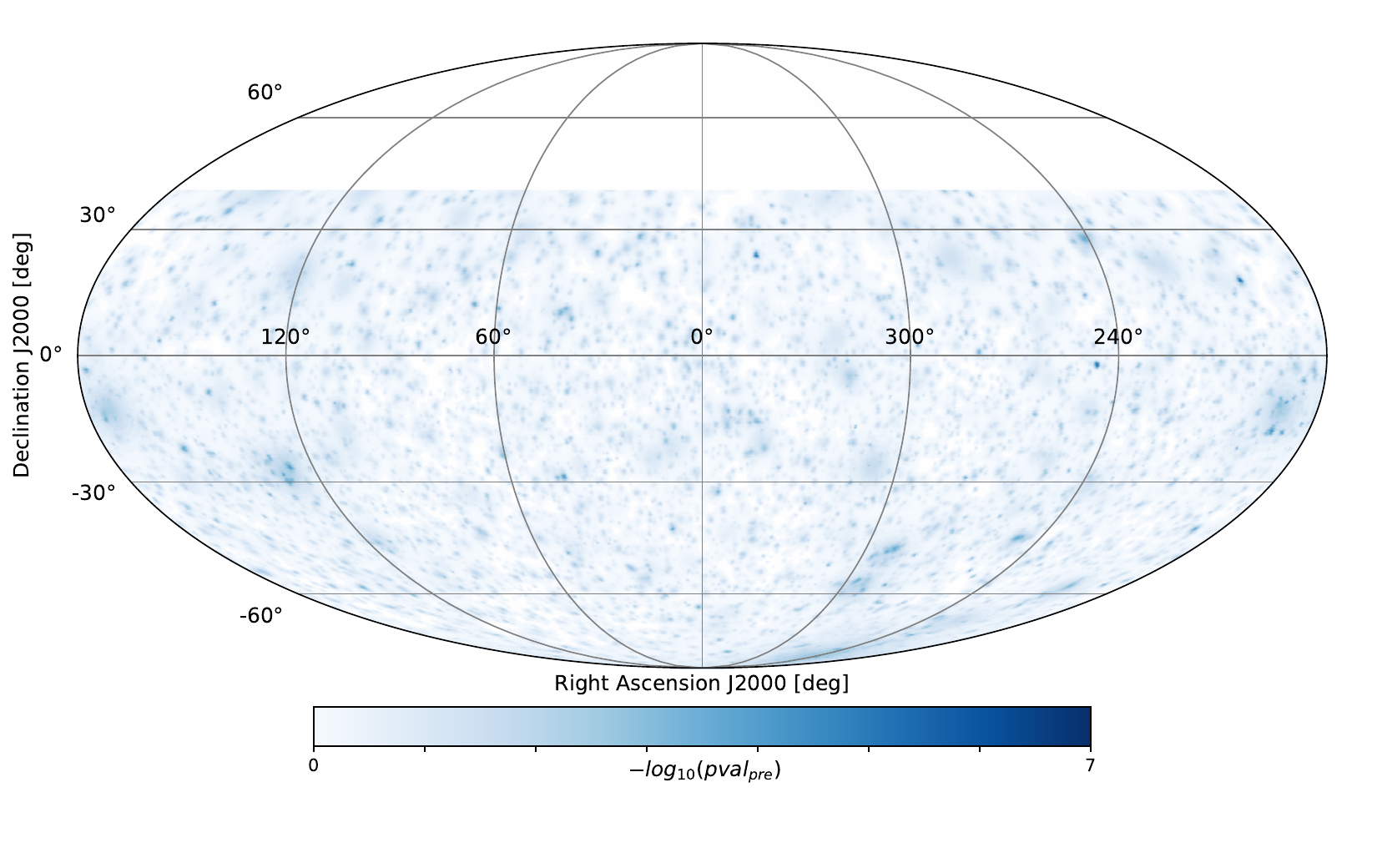}
        \put (88.5,41.3){\Large\textcolor{blue}{$\Box$}}
    \end{overpic}
    \caption{Sky map in equatorial coordinates of pre-trial p-values found in the search for point-like neutrino sources over the full ANTARES visible sky. The blue box indicates the location with the highest significance.}
\label{fig:MapFSS}
\end{figure*}

The result of the search in terms of pre-trial p-value sky map is shown in Figure~\ref{fig:MapFSS}. The most significant spot is found from the direction ($\alpha$, $\delta$) = (200.5$^\circ$\!, 17.7$^\circ$) with a best-fit number of signal events of 2.2 and a pre-trial p-value of 3.2~$\times \; 10^{-6}$, corresponding to 4.5$\sigma$. The significance in units of standard deviations is calculated adopting the one-sided Gaussian tail convention, which is consistently applied throughout the paper.

\begin{figure*}[ht!]
\centering
    \includegraphics[width=0.42\textwidth]{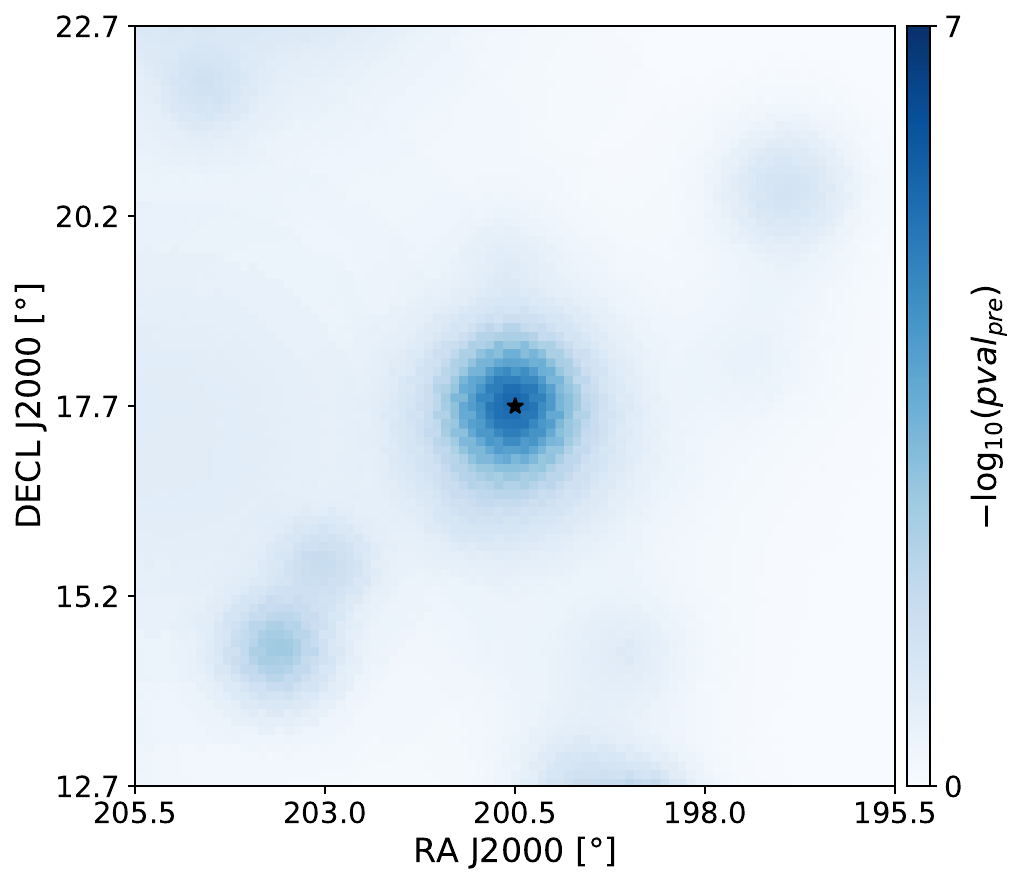}
    \includegraphics[width=0.5\textwidth]{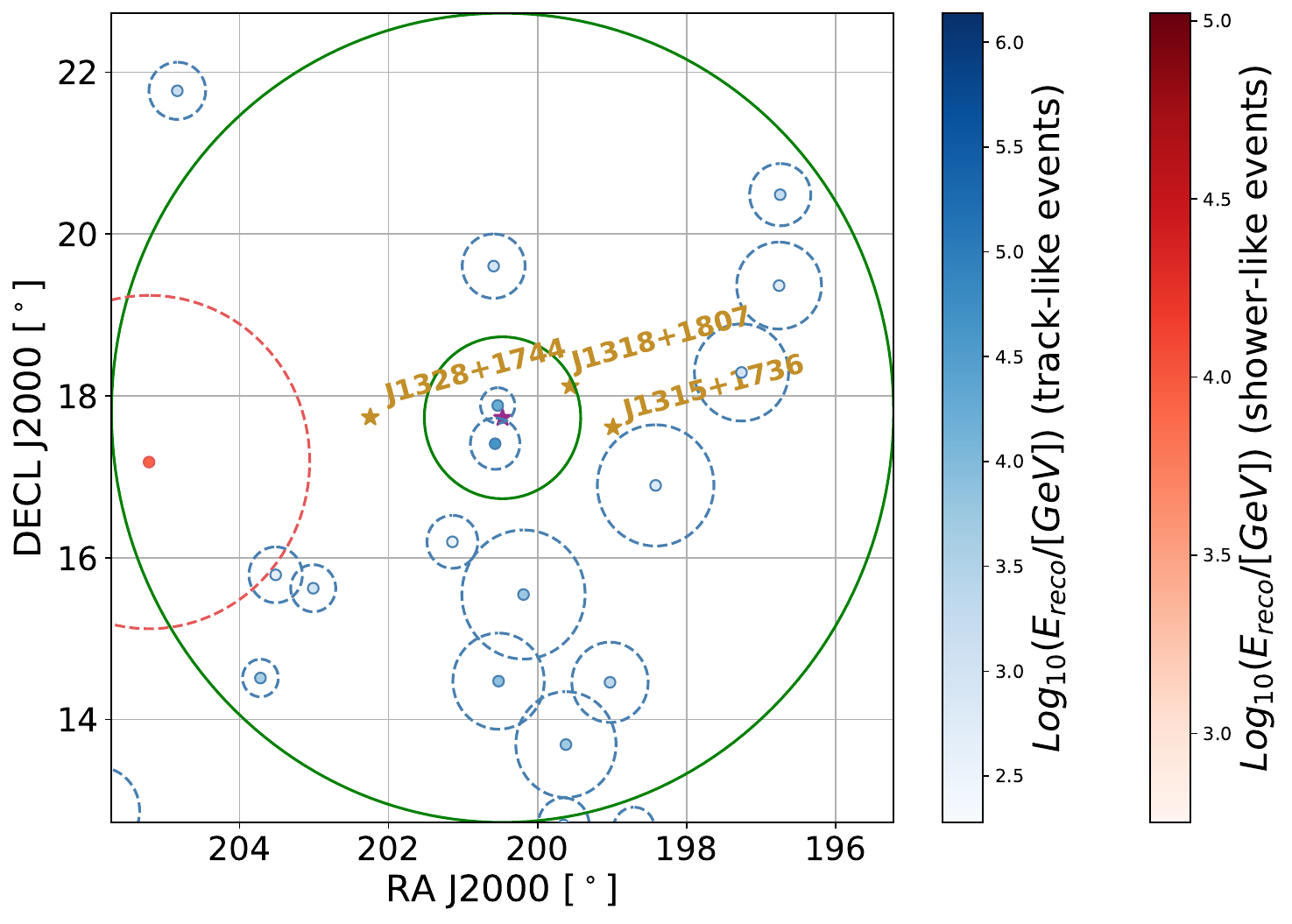}
\caption{Left: Pre-trial p-value map around the location of the full-sky most significant excess. Right: Distribution of the ANTARES events close to this region. The inner (outer) solid green line depicts the one (five) degree distance from the position of the hotspot. The red points denote shower-like events, whereas the blue points indicate track-like events. The dashed circles around the events indicate the angular error estimate. Different tones of red and blue correspond to the values assumed by the energy estimators as shown in the legend. The location and names of the three astrophysical sources located closer than 2.0$^\circ$ from the hotspot are shown in golden.}
\label{fig:hotspotevents}
\end{figure*}

Figure~\ref{fig:hotspotevents}-left shows the pre-trial p-values sky map in a $10^{\circ}\, \times \, 10^{\circ}$ square centred at the location of the hotspot, while Figure~\ref{fig:hotspotevents}-right depicts the ANTARES events reconstructed close to the same sky direction. The closest astrophysical object is blazar {J1318+1807}, located at a distance of $\sim$$1^\circ$ from the hotspot. Other sources within $2.0^\circ$ are {J1315+1736} ($\sim$$1.5^\circ$) and {J1328+1744} ($\sim$$1.8^\circ$).
For the full-sky hotspot, a post-trial p-value of 38\% is obtained.

\subsection{Galactic Plane scan} 
\label{subsec:gal-plane}

In this approach, the same analytical method used in the time-integrated full-sky search is applied to a restricted sky region containing the Galactic Plane in search of extended neutrino emission. Recent gamma-ray observations from instruments like HAWC~\cite{HAWC:2020hrt} and LHAASO~\cite{LHAASO:2023rpg} have provided compelling evidence for multiple Galactic sources displaying an extended spatial morphology, which motivates the search for extended neutrino sources. Such objects show energy spectra continuing beyond 100 TeV, which is a promising signature of hadronic interactions and therefore hint to high-energy neutrino emission. In fact, the IceCube Collaboration has reported the presence of a neutrino hotspot compatible with a source with an extension of $1.7^\circ$ at $2.6\sigma$ significance near source \hbox{3HWC J1951+266}~\cite{IceCube:2023ujd}. Moreover, the measured excess from the Galactic Plane~\cite{IceCube:2023ame}, also hinted by ANTARES data~\cite{ANTARES:2022izu}, does not exclude the contribution of unresolved sources. The ANTARES detector, benefiting from its clear view of the Southern Sky, becomes a valuable tool for these studies.

The targeted region is a rectangular area comprised within the Galactic latitude $|b|< 5^\circ$ and longitude $|l|<180^\circ$, although a portion of this region, i.e.\, most of the directions in the Galactic longitude range between $70^\circ$ and $170^\circ$, is not visible since it does not produce upgoing events in the ANTARES detector.
The region is divided into a grid of the same resolution as the one employed in the full-sky search. 
Assuming that the centre of each pixel is the centre of a neutrino extended source, the data are inspected to look for the presence of such a source, testing different Gaussian widths $\sigma_\text{Ext}$: $0.5^\circ$, $1.0^\circ$, $1.5^\circ$, and $2.0^\circ$. 

\begin{figure} [htb!]
    \centering
    \includegraphics[width=\linewidth]{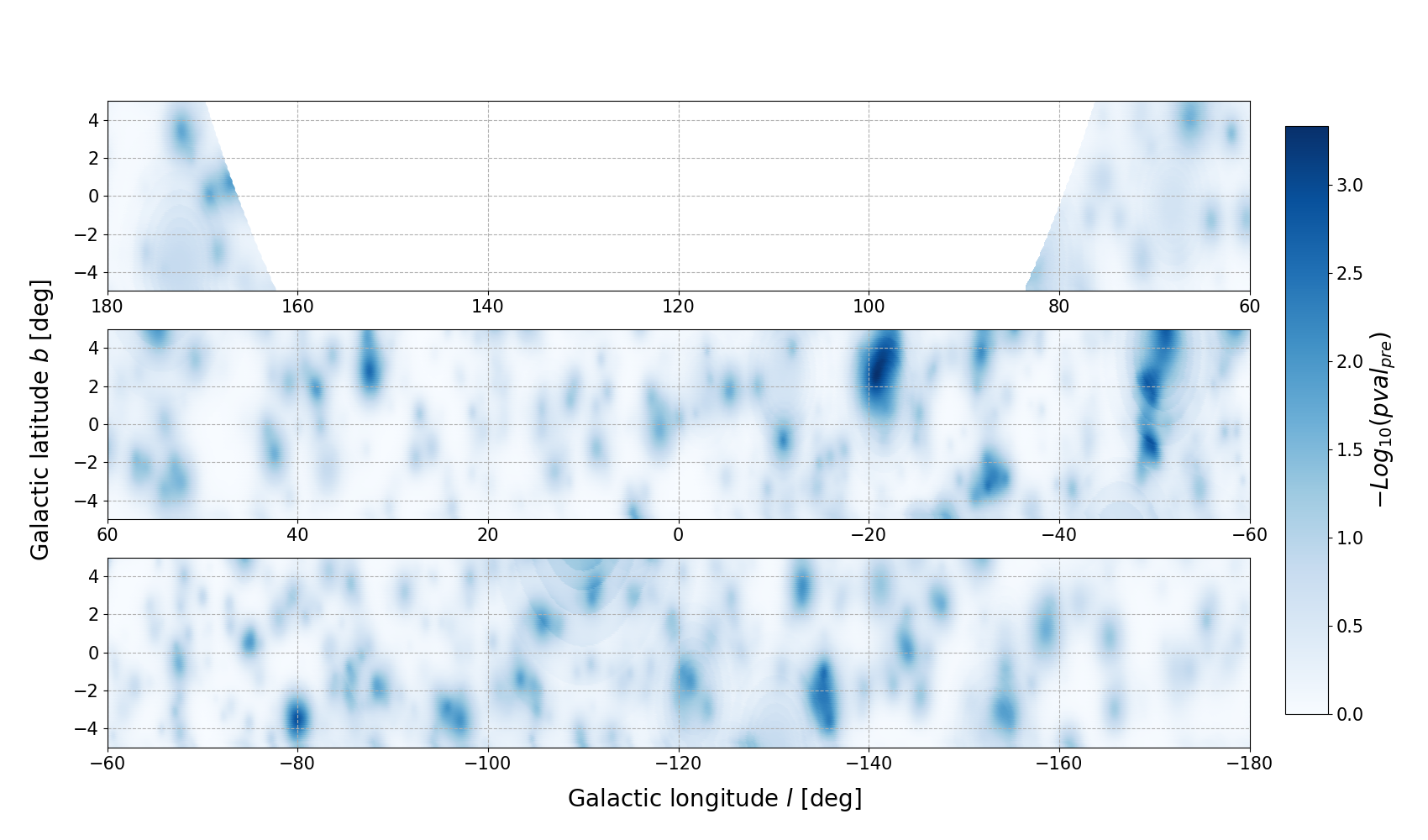}
    \caption{Colour map of the obtained pre-trial p-values at each of the analysed locations of the Galactic Plane. The presence of a source with a Gaussian width of $1.0^\circ$ has been assumed in the likelihood. Most of the directions between $70^\circ$ and $170^\circ$ in the Galactic longitude are not accessible as they produce only downgoing events in the ANTARES detector.}
    \label{fig:Legacy_GalPlane}
\end{figure}

For all the tested extensions, the hotspot is found at the same Galactic coordinates $(l,b)=(-20.7^\circ,2.6^\circ)$, with the lowest pre-trial p-value being $3.0\times10^{-4}$ found for the case of $\sigma_\text{Ext}=0.5^\circ$. When each of the best local p-values is corrected for the number of trials, the best post-trial p-value is found instead for $\sigma_\text{Ext}=1.0^\circ$.  However, the resulting significance remains very low, with the hotspot being 60\% compatible with the background. Since the obtained result is not significant, no additional correction for the number of tested extension sizes has been applied, nor has any search for associated source been performed.
The pre-trial p-value sky map of the Galactic Plane for $\sigma_\text{Ext}=1.0^\circ$ is shown in Figure~\ref{fig:Legacy_GalPlane}.

\subsection{Candidate list search}
\label{subsec:candidate-list}

In this search, the trial factor penalization that comes from inspecting many different sky locations is drastically reduced by investigating only a pre-selected list of promising cosmic neutrino sources. The list of targeted objects updates the one analysed in~\cite{Illuminati:2023}, which contained neutrino source candidates both from Galactic and extragalactic origin mostly selected from the TeVCat catalogue~\cite{Wakely:2007qpa}, the first LHAASO catalogue~\cite{LHAASO:2023rpg} and the third HAWC catalogue~\cite{ANTARES:2020zng}. Among all the changes, the following ones are the most remarkable. Only the LHAASO sources detected by the high-energy telescope KM2A have been kept and the recently detected \textit{super-PeVatron}, Cygnus OB2 bubble, has been added \cite{LHAASO:2023uhj}. The stellar clusters Danks~1, Westerlund~1 and NGC~3603 have been added as potential Galactic PeVatrons~\cite{Mitchell:2024yuy} together with the star-forming galaxies SMC, Circinus, NGC~4945 and ARP from~\cite{Ajello:2020zna}. 
 Finally, three sources previously highlighted in ANTARES studies have been added to the list for further monitoring: 
3C403~\cite{ANTARES:2020zng}, MG3~J225517+2409~\cite{ANTARES:2020zng}, and J0242+1101~\cite{ANTARES:2023lck}. 

All the sources that have a reported extension in gamma-ray emission  larger than $0.4^\circ$ have been considered as extended during the likelihood evaluation.

A total of 161 point-like and eight extended source candidates have been inspected.   Their name and equatorial coordinates, together with their spatial extension, if applicable, are listed in Table~\ref{tab:LimitsFix} -- only for the six most significant point-like objects -- and Table~\ref{tab:LimitsFixExt} for all the extended sources. The information about all the analysed sources can be found in the Appendix~\ref{Appendix}.
The tables also report, for each object, the results found in the search in terms of best-fit number of signal events ($\hat{\mu}_\mathrm{sig}$), pre-trial p-value (p-val), and 90\% confidence level (CL) upper limits on the one-flavour flux normalisation factor ($\Phi_{1\rm\, GeV}^{90\%}$), for $\gamma = 2.0$ and $\gamma = 2.5$ resulting from the parametrisation shown in Eq. \eqref{eq:FluxParam} in units of ${\rm 10^{-8} \, (10^{-6}) \, \times GeV^{-1}cm^{-2}s^{-1}}$. 

\begin{figure*}[ht!]
\centering
    \includegraphics[width=0.42\textwidth]{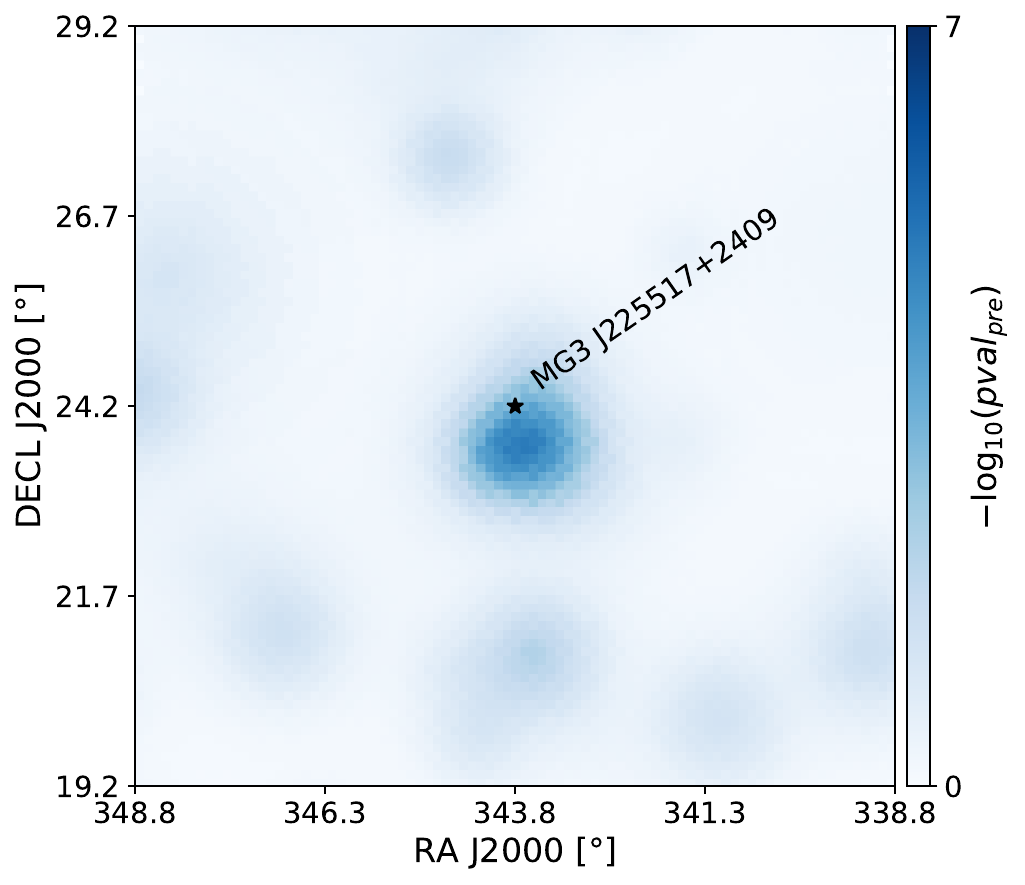}
    \includegraphics[width=0.5\textwidth]{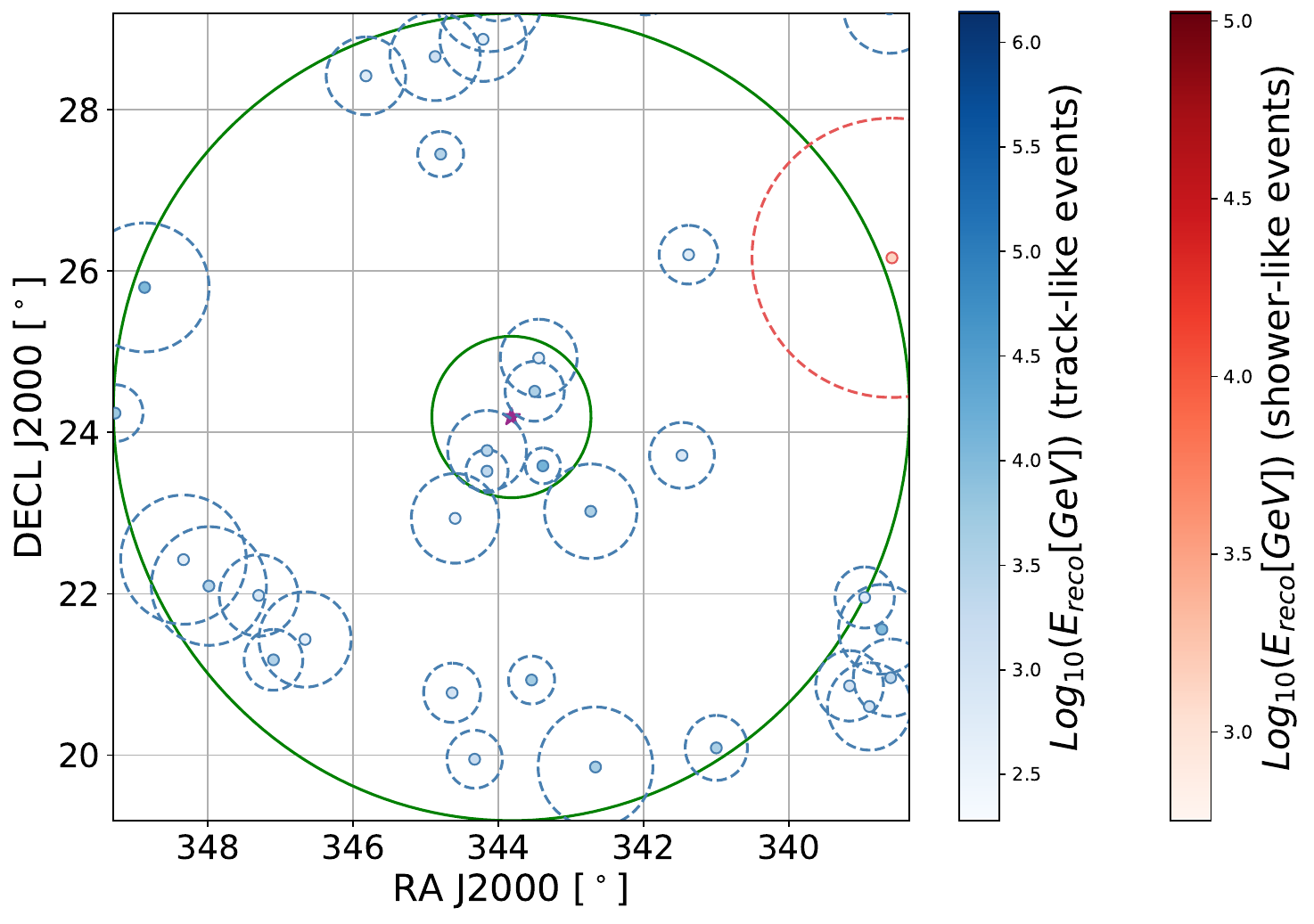}
\caption{Left: Pre-trial p-value map around the location of the most significant source MG3 J225517+2409 for $\gamma = 2.0$. Right: Distribution of the ANTARES events close to MG3 J225517+2409. A detailed description of the map is given in the caption of Figure \ref{fig:hotspotevents}.
}
\label{fig:Maps-CL}
\end{figure*}

The most significant candidate for both values of $\gamma$ is MG3~J225517+2409, with a pre-trial p-value of  $2.4\times10^{-4}\,(3.5\sigma)$ for $\gamma = 2.0$ and $6.4\times10^{-5}\,(3.8\sigma)$ for $\gamma = 2.5$. The pre-trial p-value sky map around the location of blazar MG3~J225517+2409 is shown in Figure~\ref{fig:Maps-CL}-left for $\gamma = 2.0$, while Figure~\ref{fig:Maps-CL}-right depicts the ANTARES events located close to the same source.  
MG3J225517+2409 was included in the candidate list prompted by an excess observed in a previous ANTARES search~\cite{ANTARES:2020zng}. In that analysis, based on 11 years of ANTARES data, the source was an outlier among the 1255 blazars examined, with a pre-trial p-value of $1.4 \times 10^{-4}$ (post-trial p-value of 0.15), located in the extreme tail of the p-value distribution. 
In the present work, the local significance at the position of MG3J225517+2409 should be corrected for both the size of the tested source list and the \textit{a~posteriori} inclusion of this target.
Correcting only for the catalogue size reduces the significance to $2.0\sigma$. Because a robust treatment of the combined trial factor is non-trivial and the current excess is modest, no additional correction is applied. The source is therefore not considered significant.

{ 
\footnotesize
\setlength{\tabcolsep}{.3em}

\begin{longtable}{lcccccccccc}
    \caption{ List of the six most significant candidates in terms of pre-trial p-value assuming a point-like emission profile ordered from lower to higher declination.} \label{tab:LimitsFix} \\
    
    \multirow{3}{*}{source name} & 
    \multirow{3}{*}{$\delta [^{\circ}]$} &
    \multirow{3}{*}{$\alpha [^{\circ}]$} & &
    \multicolumn{3}{c}{$\gamma = 2.0$} & ~ &
    \multicolumn{3}{c}{$\gamma = 2.5$} \\ \cline{5-7}\cline{9-11}\\[-6pt]
        &  &  & &
    $\hat{\mu}_\mathrm{sig}$ & p-val & $\Phi_{1\rm\, GeV}^{90\%}$ & ~ &
    $\hat{\mu}_\mathrm{sig}$ & p-val & $\Phi_{1\rm\, GeV}^{90\%}$ \\[2pt] \hline \\[-6pt]

    Galactic Centre & $-29.01$ & 266.43 & ~ & 2.1 & 0.017 & 1.2 & ~ & 2.4 & 0.012 & 2.8 \\
    J0609-1542 & $-20.12$ & 287.79 & ~ & 1.2 & 0.0073 & 1.4 & ~ & 1.3 & 0.018 & 2.9 \\
    3C403 & 2.51 & 298.07 & ~ & 2.5 & 0.00048 & 2.0 & ~ & 2.6 & 0.00047 & 5.3 \\
    TXS 0506+056 & 5.69 & 77.35 & ~ & 2.2 & 0.0075 & 1.6 & ~ & 2.6 & 0.0087 & 4.0 \\
    J0242+1101 & 11.02 & 40.64 & ~ & 3.7 & 0.0074 & 1.6 & ~ & 5.3 & 0.0039 & 4.6 \\ 
    MG3 J225517+2409 & 24.19 & 343.82 & ~ & 4.0 & 0.00024 & 2.3 & ~ &4.8 & 0.000064 & 8.2 \\ 
    
    \hline
    
\end{longtable}

}

{ 
\footnotesize
\setlength{\tabcolsep}{.3em}

\begin{longtable}{lcccccccccc}
    \caption{List of the eight source candidates analysed assuming an extended emission profile ordered from lower to higher declination. Dashes (--) in the fitted number of source events and pre-trial p-value indicate sources with null fitted signal.} \label{tab:LimitsFixExt} \\
    \multirow{3}{*}{source name} & 
    \multirow{3}{*}{$\delta [^{\circ}]$} &
    \multirow{3}{*}{$\alpha [^{\circ}]$} & 
    \multirow{3}{*}{$\sigma_\text{Ext} [^{\circ}]$} & 
    \multicolumn{3}{c}{$\gamma = 2.0$} & ~ &
    \multicolumn{3}{c}{$\gamma = 2.5$} \\ \cline{5-7}\cline{9-11}\\[-6pt]
    &  &  &  & 
    $\hat{\mu}_\mathrm{sig}$ & p-val & $\Phi_{1\rm\, GeV}^{90\%}$ & ~ &
    $\hat{\mu}_\mathrm{sig}$ & p-val & $\Phi_{1\rm\, GeV}^{90\%}$ \\[2pt] \hline \\[-6pt]

    Danks 1 & $-62.69$ & 198.12 & 0.66 &  -- & -- & 0.43 & ~ & -- & -- & 0.78 \\
    RX J0852.0-4622 & $-46.37$ & 133.00 & 0.63 & -- & -- & 0.45 & ~ & -- & -- & 0.89 \\
    Westerlund 1 & $-45.85$ & 251.76 & 0.80 &  1.5 & 0.11 & 0.79 & ~ & 2.0 & 0.14 & 1.84 \\ 
    Vela X & $-45.60$ & 128.75 & 0.58 & 2.8 & 0.027 & 0.97 & ~ & 3.2 & 0.040 & 2.2 \\
    1LHAASO J1928+1813u & 18.23 & 292.07 & 0.63 &  -- & -- & 1.0 & ~ & 0.27 & 0.28 & 2.4 \\
    3HWC J1951+266 & 26.61 & 297.90 & 1.7 &  -- & -- & 0.93 & ~ & -- & -- & 2.9 \\
    1LHAASO J2028+3352 & 33.88 & 307.21 & 1.70 &  --  & -- & 0.52 & ~ & -- &-- & 1.8 \\
    CygOB2 & 41.04 & 307.44 & 2.17 & -- & -- & 0.59 & ~ & -- & --  & 2.4 \\ \hline
    
\end{longtable}

}

\begin{figure*}[ht!]
\centering
    \begin{subfigure}[b]{0.41\textwidth}
         \centering
         \begin{overpic}[width=\linewidth]{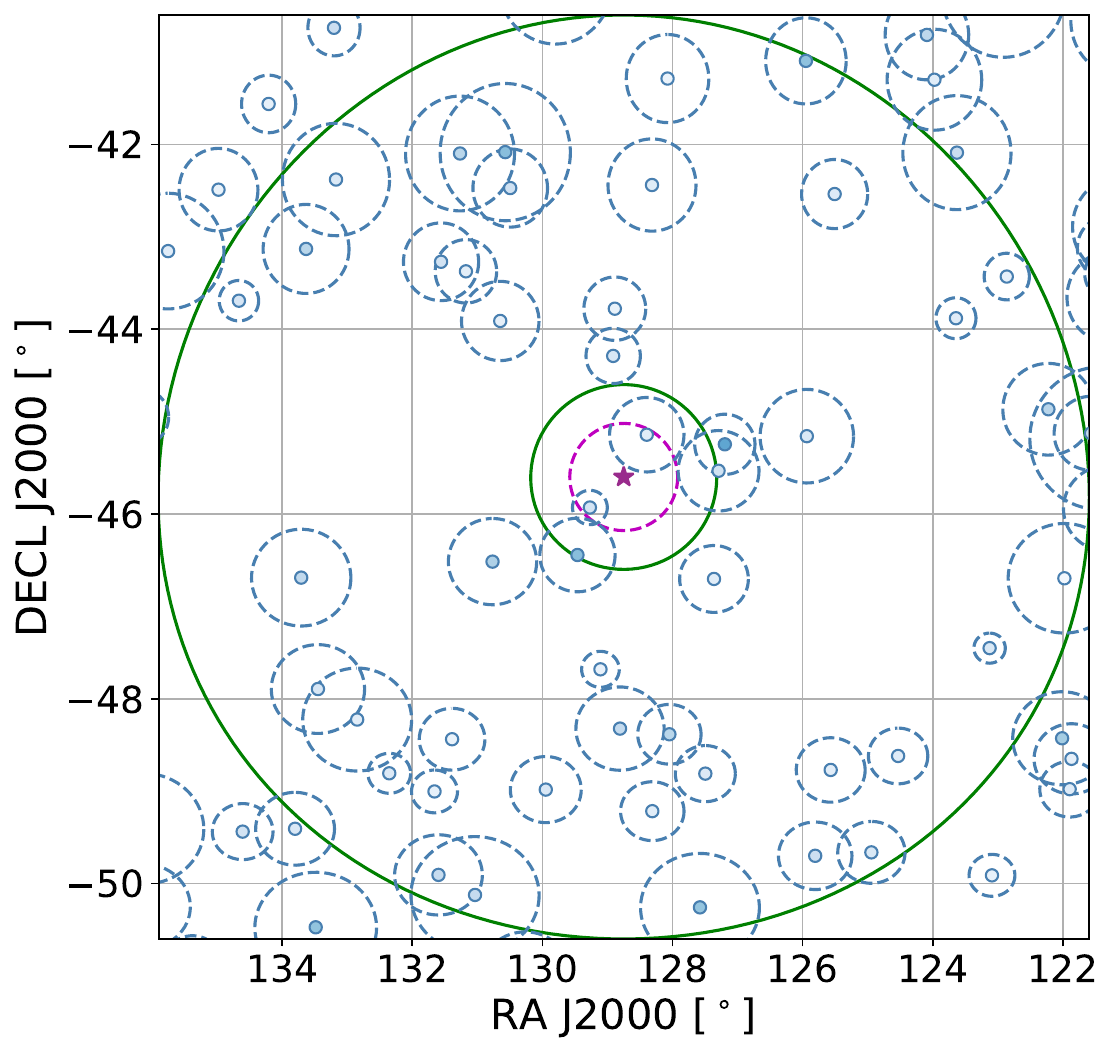}
        \end{overpic}
         \caption{}
         \label{fig:VelaX_scatter}
    \end{subfigure}
    \vspace{-0.3cm}
    \begin{subfigure}[b]{0.555\textwidth}
         \centering
         \begin{overpic}[width=1.\linewidth]{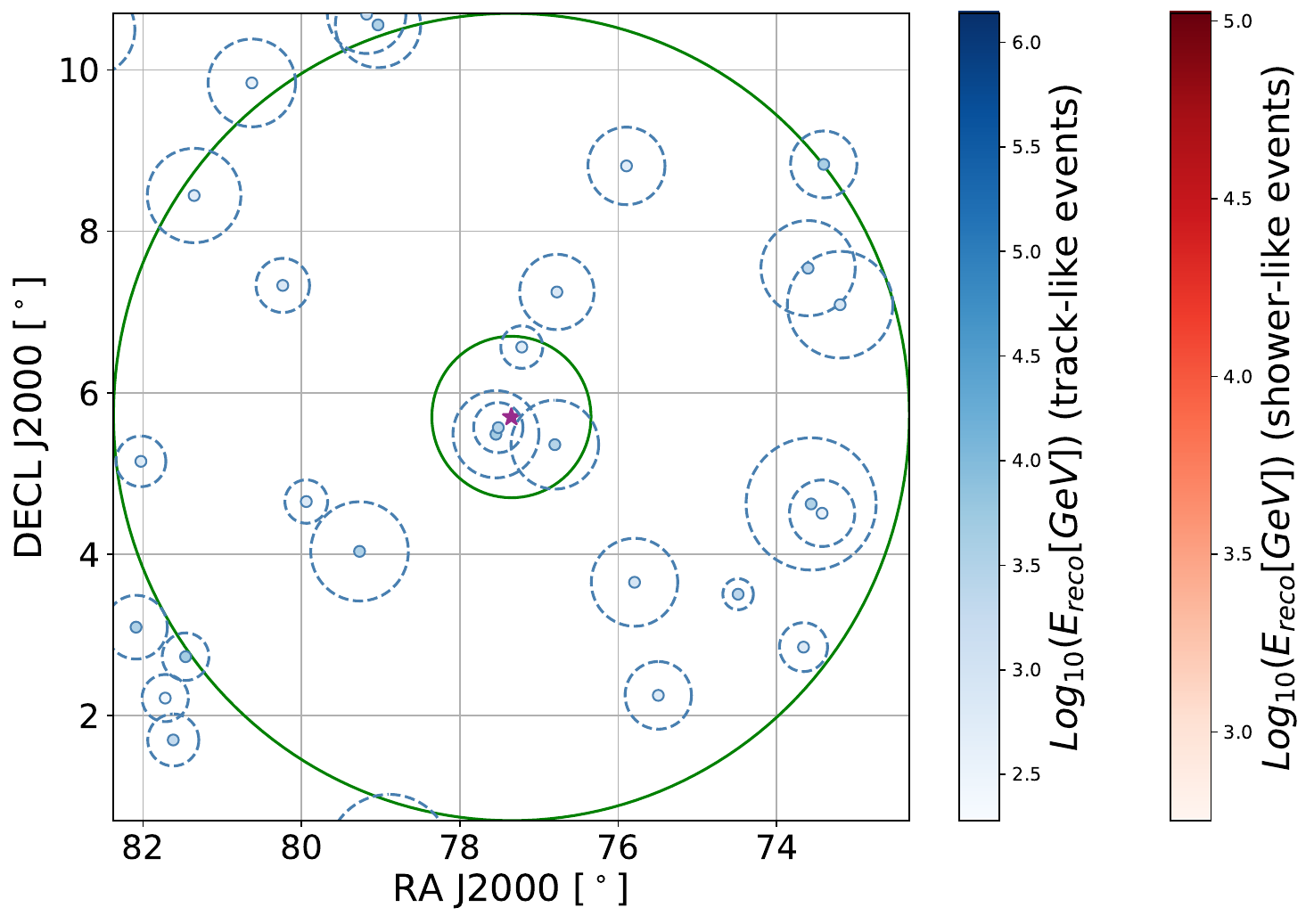}
         \end{overpic}
         \caption{}
         \label{fig:TXS_Scatter}
    \end{subfigure}
\caption{Distribution of the ANTARES events close to \textbf{(a)} Vela~X, with the magenta dashed line indicating the Gaussian extension assumed in the analysis, and to \textbf{(b)} TXS~0506+056. A detailed description of the map is given in the caption of Figure \ref{fig:hotspotevents}.
}
\label{fig:Maps-CL2}
\end{figure*}

In the case of the extended source candidates, it is highlighted that the pulsar wind nebula Vela~X has seen its significance increased from $1.3\sigma$ ($\hat{\mu}_\text{sig} = 1.9$)~\cite{Illuminati:2023} to almost $2.0\sigma$ ($\hat{\mu}_\text{sig} = 2.8$) after an extension of $0.58^\circ$ was assumed. As it can be seen in Figure \ref{fig:Maps-CL2}-left, the likelihood can better accommodate the two high-energy track-like events within the $1^\circ$ distance. 

\begin{figure*}[ht!]
\centering
\includegraphics[width=1.02\linewidth]{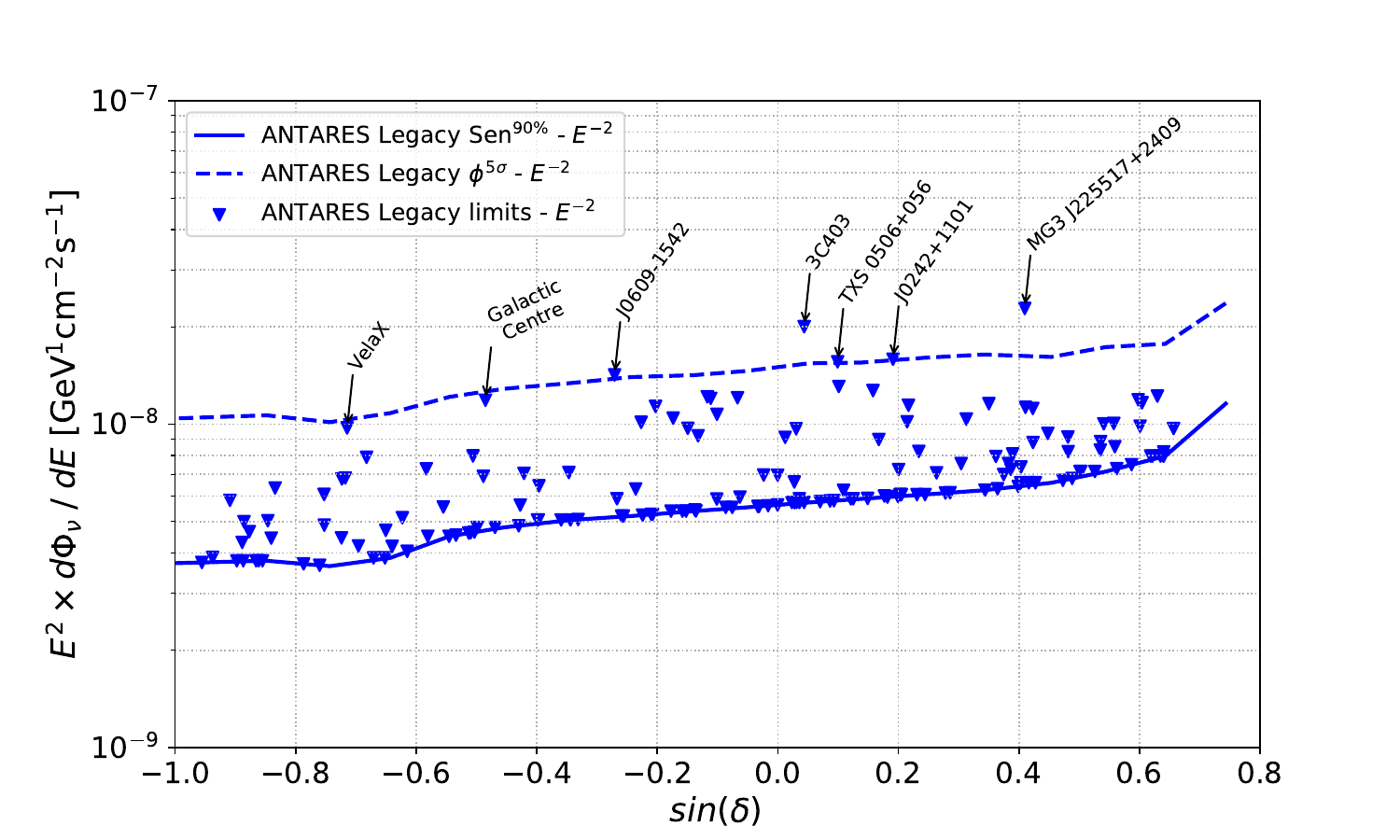}
\includegraphics[width=1.02\linewidth]{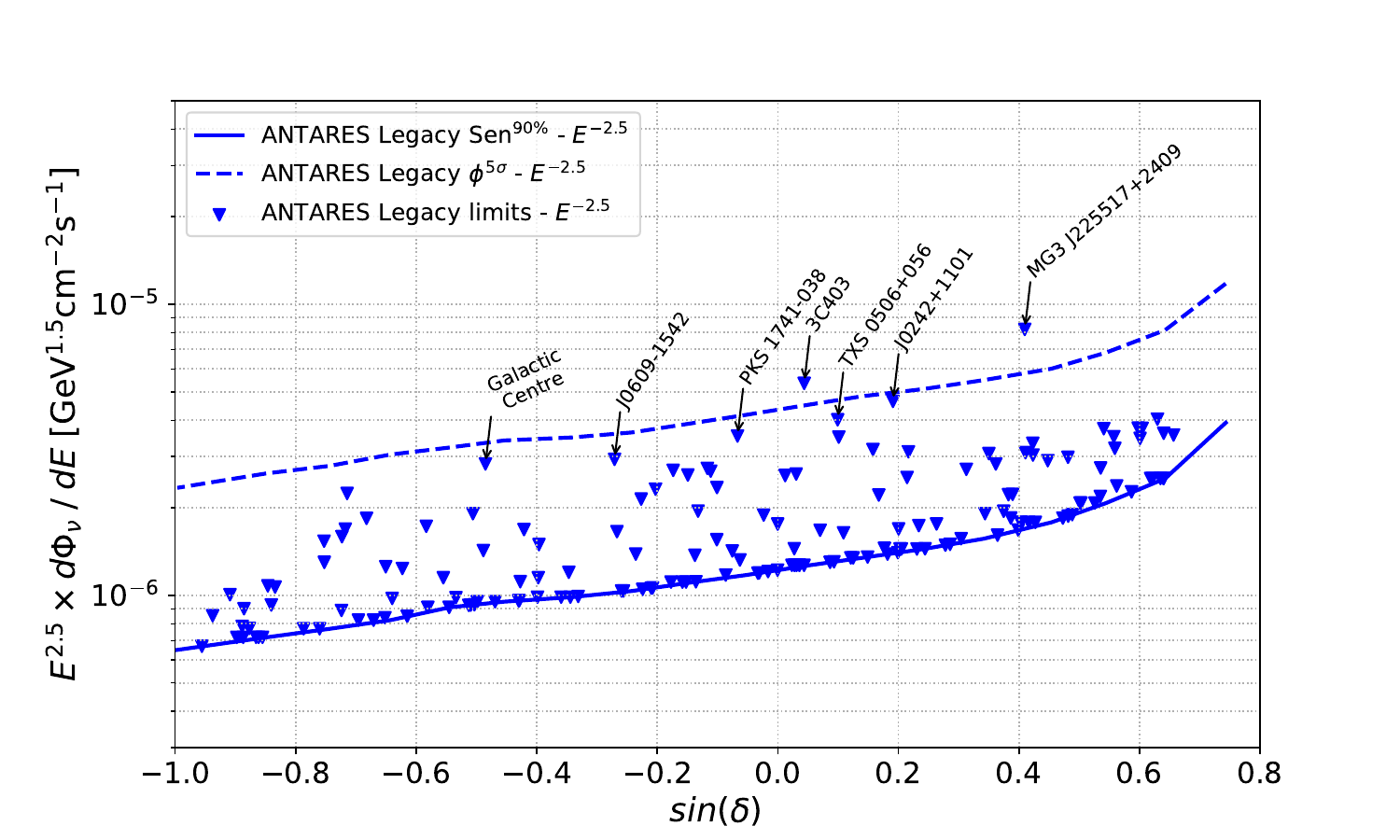}
\caption{Upper limits at $90\%$~CL (triangles) on the one-flavour neutrino flux normalisation for the investigated astrophysical candidates as a function of the source declination. The top panel shows results for an $E^{-2.0}$ energy spectrum while the bottom corresponds to an $E^{-2.5}$ spectrum. The arrows point to the the upper limits corresponding to the sources for which a pre-trial significance over $2.0\sigma$ has been obtained. The solid (dashed) line shows the $90\%$~CL median sensitivity ($50\%$ $5\sigma$~discovery flux) of the analysis.}
\label{fig:Limits}
\end{figure*}

The 90\% CL upper limits on the one-flavour neutrino flux normalisation for the 169 investigated candidates are shown in Figure~\ref{fig:Limits} as a function of the declination. The $5\sigma$ discovery potential of the analysis, defined as the neutrino flux needed for a $5\sigma$ discovery in 50\% of the trials, and the 90\% CL\ sensitivity, i.e.\, the median expected 90\% CL upper limit on the flux normalisation in case of pure background, are also shown.

The analysed list also includes the most promising sources indicated by the latest IceCube results. Blazar TXS 0506+056 has been constantly monitored since the first evidence of neutrino emission was observed. The first time-integrated search  using 11 years of data fitted 1.03 signal events around the blazar with a local significance of $2.1\sigma$ \cite{Aublin:2019zzn}. In the 13-years update, the significance of TXS 0506+056 raised to $2.8\sigma$ due to the detection of 2 additional events near the source $(\hat{\mu}_\text{sig} =2.9)$~\cite{Illuminati:2021}. After the recent recalibration, the position of one of the close-by events was shifted away from TXS~0506+056, decreasing its significance to the value here presented: $2.4\sigma$ with 2.2 fitted signal events, shown in Figure \ref{fig:Maps-CL2}-right. The active galactic nuclei NGC~1068 has also been considered in the list. Since the results of the IceCube Collaboration point toward a much softer spectrum, $\gamma =3.2$~\cite{IceCube:2022der}, than the ones considered in this work, a dedicated search was performed assuming $\gamma=3.0$ instead. A total of 0.8 signal events were fitted by the likelihood with a local p-value of 0.22. Upper limits at the 90\% CL have been computed on the flux normalisation defined in Eq.~\eqref{eq:FluxParam}: $\Phi_{\rm 1\,TeV}^{\nu + \bar{\nu}} = 1.9\times10^{-10} \; \rm{TeV^{-1}cm^{-2}s^{-1}}$ ($\Phi_{\rm 1\,GeV}^{\nu + \bar{\nu}} = 1.9\times10^{-4} \; \rm{GeV^{-1}cm^{-2}s^{-1}}$), compatible with the IceCube observation. 

Among the sources of interest, blazar PKS~1424+240, the second most significant IceCube source~\cite{IceCube:2022der}, and the X-ray bright Seyfert Galaxies NGC~4151 and CGCG~420-015~\cite{IceCube:2024dou} are also included. No signal is observed for any of the three directions and flux models, so upper limits are reported.

\section{Results of the time-dependent searches}
\label{sec:TimeDepResults}

\begin{figure*}[ht!]
    \includegraphics[width=\textwidth]{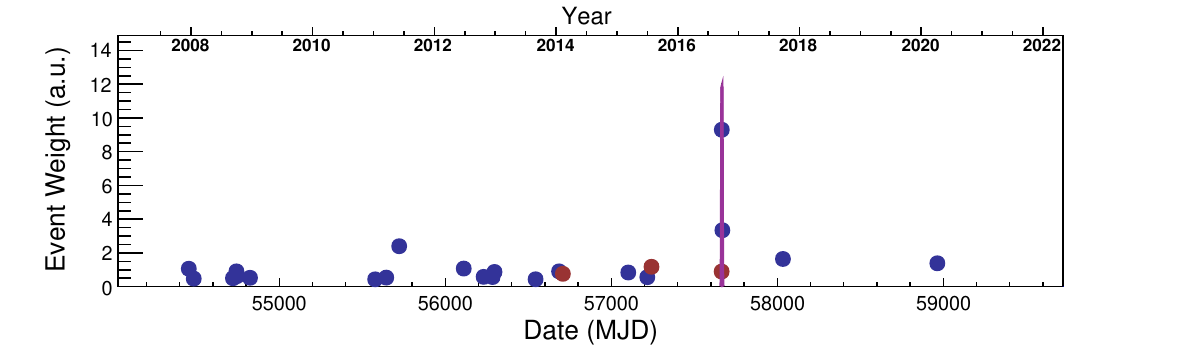}\\
    \vspace{0.00mm}
    \includegraphics[width=\textwidth]{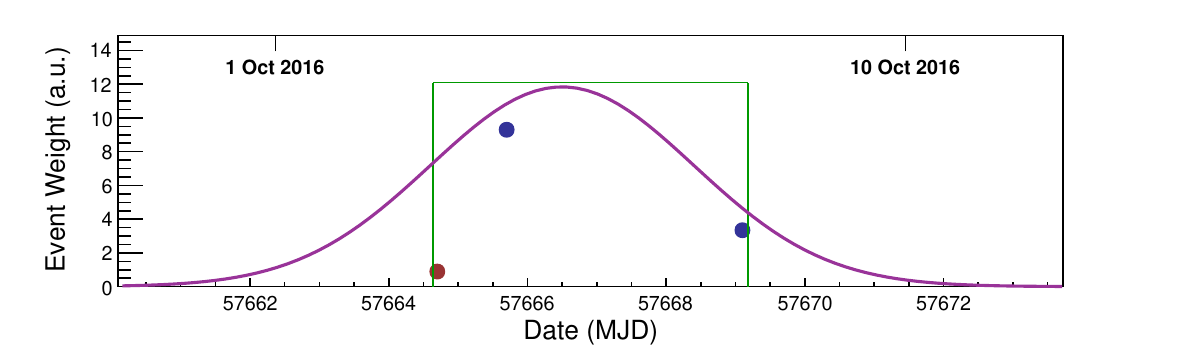}
    \caption{Weighted time distribution of ANTARES events near the hotspot of the full-sky time-dependent scan. The top panel shows events over the full livetime, while the bottom panel focuses on the period around the flare. Only track-like events within $5^\circ$ (shown in blue) and shower-like events within $10^\circ$ (shown in red) of the hotspot are included in the plot. A higher weight is associated to events with smaller distance to the source and larger value of the energy estimator.
    The magenta and green lines represent the best-fit Gaussian and box profiles. The hotspot location used for this plot corresponds to that of the best-fit Gaussian flare, which differs by $0.4^\circ$ from the box-profile flare location.}
    \label{fig:TimeDistrHotspot}
\end{figure*}

Each of the sky directions defined in Section~\ref{subsec:time-int-full-sky} has also been tested for time-dependent neutrino emissions. 
The most significant Gaussian-shaped flare is found at the location \hbox{($\alpha$, $\delta$) = (141.3$^\circ$\!, 9.8$^\circ$)}, with the following best-fit parameters from the maximisation: \hbox{$(\hat{\mu}_\text{sig}, \hat{\gamma}, \hat{T}_0 \textrm{[MJD]}, \hat{\sigma}_t  \textrm{[days]}) = (3.0, 2.4, 57666.5, 1.9)$}. The local \hbox{p-value} of the flare is of $8.3 \times 10^{-6}$ (4.3$\sigma$),  which results into a post-trial p-value of 30\%.  
Assuming a box-shaped time profile, the most significant flare is found at coordinates \hbox{($\alpha$, $\delta$) = (141.1$^\circ$\!, 9.4$^\circ$)}, which is consistent with the direction of the best-fit flare obtained under the Gaussian shape hypothesis, considering the angular resolution of the dataset. For the box-shaped flare, the likelihood maximum is found at \hbox{$(\hat{\mu}_\text{sig}, \hat{\gamma}, \hat{T}_0 \textrm{[MJD]}, \hat{\sigma}_t  \textrm{[days]}) = (3.2, 2.4, 57666.9, 2.3)$}. A pre-trial p-value of $8.8 \times 10^{-6}$ is observed (4.3$\sigma$),  corresponding to a post-trial p-value of 30\%.  The time distribution of the ANTARES events close to the hotspot location, together with the best-fit signal time profiles, are shown in Figure~\ref{fig:TimeDistrHotspot}, while the values of the best-fit parameters are summarised in Table~\ref{tab:untriggerhighligths}.  
The closest astrophysical source to the hotspot is the BL Lac object 5BZBJ0926+0803, located at an angular distance of 1.8$^\circ$ (1.5$^\circ$) from the hotspot position found assuming the Gaussian (box) shape hypothesis.

As an additional study, sky locations identified as interesting in previous analyses were revisited. 
Among them is the direction of the radio-bright blazar J0242+1101, also known as PKS~0239+108, for which a potential neutrino flare coincident with a radio and a $\gamma$-ray flare had been reported in a previous ANTARES study~\cite{ANTARES:2023lck}. In this analysis, the fitted flare parameters for J0242+1101 -- reported in Table ~\ref{tab:untriggerhighligths}-- are compatible with those obtained in the previous search within differences of only a few days. Although the multi-messenger temporal overlap is confirmed, it should be noticed that the p-value associated to this flare has increased, passing from $6.0\times10^{-3}$ ($2.1\times10^{-3}$) to $2.1\times10^{-2}$ ($8.2\times10^{-3}$) for the Gaussian-like (box-like) profile.

In addition to J0242+1101, the directions of the sources highlighted in the neutrino flare search performed by the IceCube Collaboration~\cite{IceCube:2021slf} have also been examined. That analysis targeted 110 flaring candidate sources, six of which-- 1ES~1959+650, PKS~1502+106, NGC~1068, TXS~0506+056, M87, and mbox{GB6~J1542+6129-- exhibited flare activity with pre-trial significances exceeding $2.0\sigma$, with the flare from TXS0506+056 corresponding to the one previously reported in~\cite{IceCube:2018cha}.
With the exception of 1ES~1959+650 and GB6~J1542+6129, which lie outside of the ANTARES accessible sky with upgoing events, the direction of the remaining four sources have been examined with the time-dependent approach. 

For PKS~1502+106 and TXS~0506+056, ANTARES also observes flare activity with pre-trial significance above $2.0\sigma$. Figure~\ref{fig:TimeDistrSources} shows the time distribution of ANTARES events in the vicinity of the two sources, along with the corresponding best-fit signal time profiles. For comparison, the best-fit flares reported by IceCube are also shown. The ANTARES best-fit parameters are summarised in Table~\ref{tab:untriggerhighligths}.

Interestingly, the ANTARES and IceCube flares for these sources exhibit a non-negligible temporal overlap. 
In order to quantify it, Gaussian flares are defined as $\hat{T}_0 \pm 3\hat{\sigma}_t$, while box flares as $\hat{T}_0 \pm \hat{\sigma}_t$.
For PKS~1502+106, a pre-trial significance exceeding $2.0\sigma$ is obtained for both shape assumptions. 
Both the observed ANTARES flares are fully contained within the IceCube flare window, resulting in an overlap of $100\%$ relative to the ANTARES flare duration, and $12\%$ ($7\%$) relative to the IceCube flare duration when using the Gaussian-like (box-like) ANTARES profile.
For TXS~0506+056, the only ANTARES flare with pre-trial significance exceeding $2.0\sigma$ is the one with a box-like time profile. The temporal overlap between the ANTARES box-shaped flare and the IceCube Gaussian flare corresponds to $54\%$ of the ANTARES flare duration and $55\%$ of the IceCube flare duration. 

In order to estimate the probability of such coincidences occurring by chance, independent sets of pseudo-experiments are performed for each source. In each set, ANTARES data are scrambled in right ascension while the source position is kept fixed. For each pseudo-experiment, the most significant flare in the direction of the source is identified and compared to the corresponding IceCube flare.

Adopting a conservative criterion, the chance probability is calculated as the fraction of pseudo-experiments in which a flare -- of either shape -- with significance $\ge 2.0\sigma$ is found in the direction of PKS~1502+106 (TXS~0506+056), and has an overlap of at least $7\%$ ($54\%$) with the IceCube flare measured with respect to either the ANTARES or IceCube flare duration. The resulting chance probability is approximately $0.3\%$ ($0.5\%$).

Finally, the fact that four sources (PKS~1502+106, TXS~0506+056, M87, and NGC~1068) were examined and none is \textit{a priori} privileged over the others, has been considered. The probability of finding, in the ANTARES scrambled data, a flare with significance $\ge 2.0\sigma$ and a temporal overlap of at least 7\% -- the smallest among those observed -- with the IceCube flare (with respect to either flare duration) in at least one among the four targeted sources, regardless of which, was estimated.  
This probability is found to be approximately 2\%. Requiring instead that such a coincidence occurs in at least two of the four sources -- as is observed in the real data -- reduces the probability to approximately 0.02\%. While it is important to note that the flares observed individually in ANTARES and IceCube are only mildly significant (at the $\sim$$2\sigma$ level), the probability of such flares overlapping in at least two independent source directions is unlikely to arise from background fluctuations alone, making the observed pattern particularly intriguing.

\begin{figure*}[ht!]
    \includegraphics[width=0.5\textwidth]{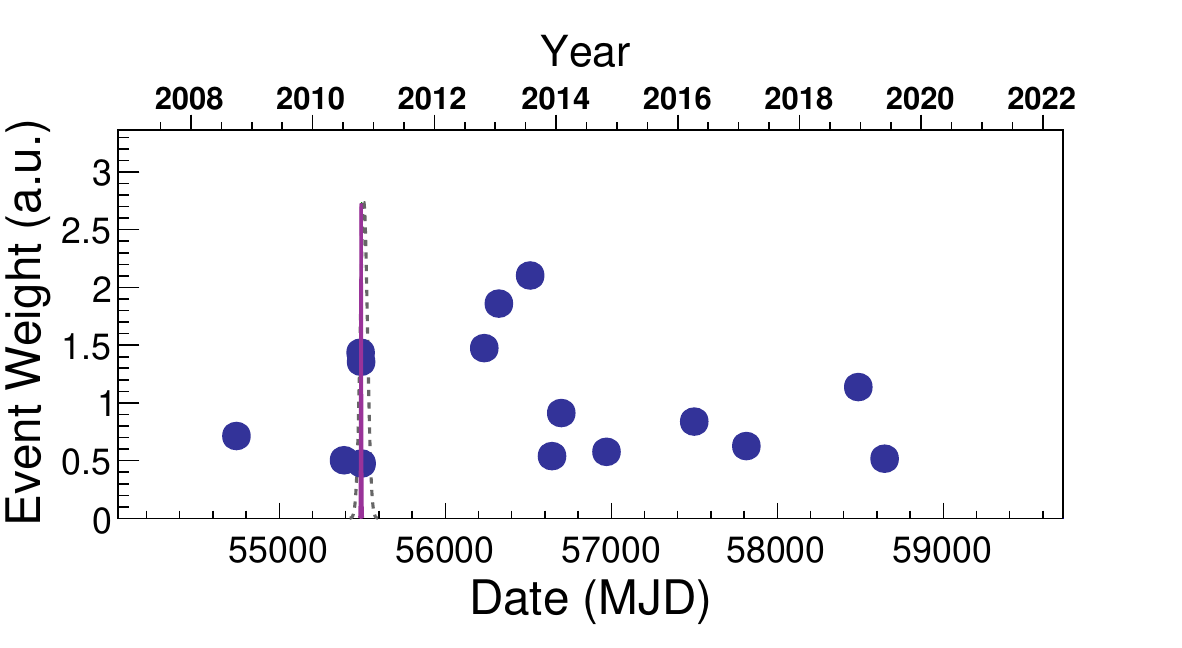}
    \includegraphics[width=0.5\textwidth]{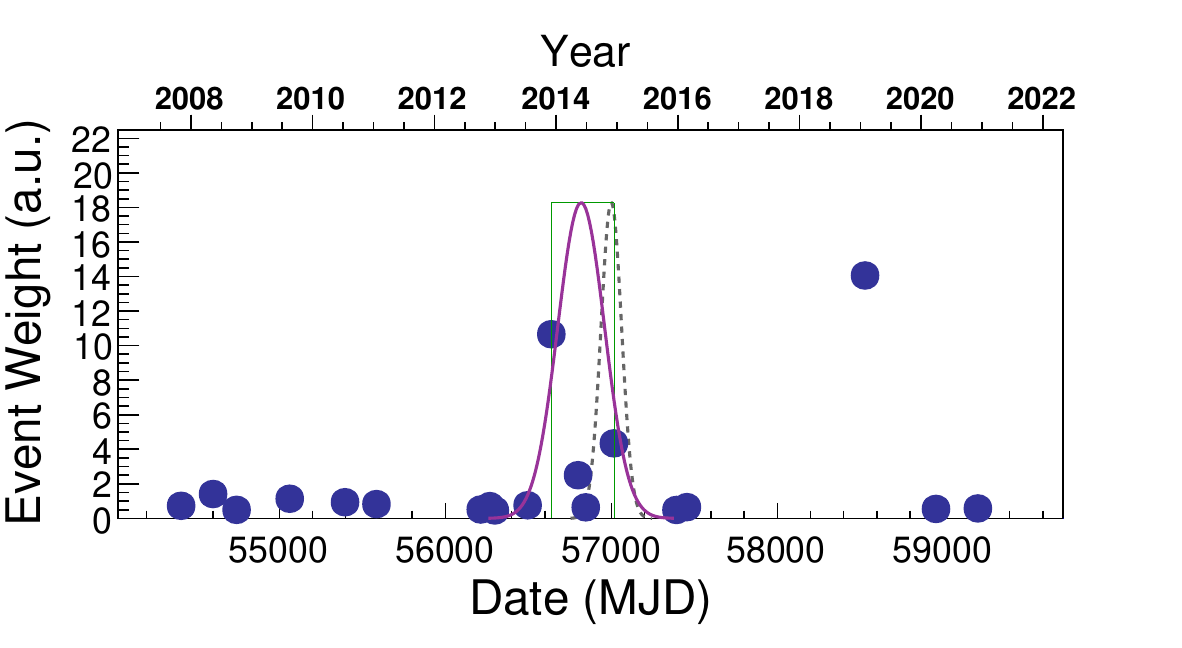}
    \includegraphics[width=0.5\textwidth]{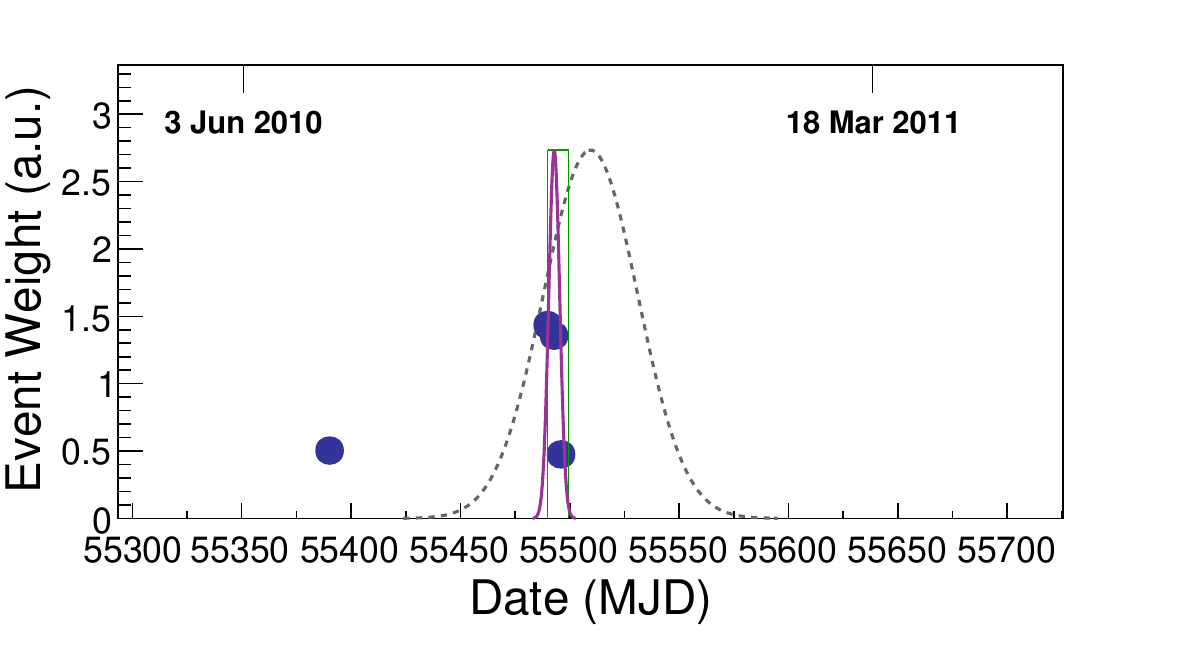}
    \includegraphics[width=0.5\textwidth]{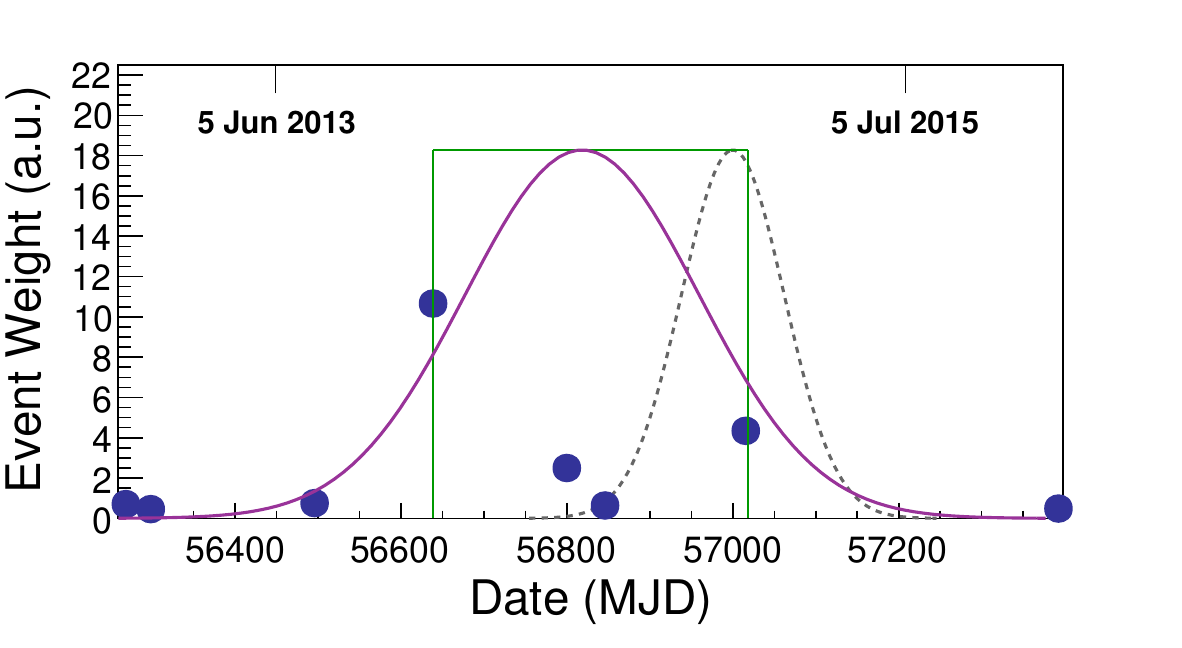}
    \caption{Weighted time distribution of ANTARES events near PKS~1502+106 (left) and TXS~0506+056 (right). The top panel shows events over the full livetime, while the bottom panel focuses on the period around the best-fit flare. Only track-like events (blue dots) within $5^\circ$ of the source location are included in the plot. No shower event was detected in the vicinity of these sky directions. A higher weight is associated to events with smaller distance to the source and larger value of the energy estimator. The magenta and green solid lines represent the best-fit Gaussian and box profiles. The grey dashed line shows the best-fit flare from the same sources identified in the study by the IceCube Collaboration~\cite{IceCube:2021slf}.}
    \label{fig:TimeDistrSources}
\end{figure*}

\renewcommand{\arraystretch}{1.2} 
\setlength{\tabcolsep}{0.25em}

\begin{table*}[ht!]
\footnotesize
\centering
\caption{
Results of the time-dependent analysis for selected directions. 
    The first two rows present the findings for the full-sky hotspot found assuming Gaussian (G) and box-shaped (B) signal time profiles, defined according to Eq.~\ref{eq:time_gauss} and~\ref{eq:time_box}. The following rows detail the results from interesting directions identified in previous studies, as detailed in the main text.
    In addition to equatorial coordinates, the table presents the best-fit parameters for each identified flare, including central time $\hat{T}_0$, flare duration $\hat{\sigma}_t$, number of signal events $\hat{\mu}_\mathrm{sig}$, spectral index $\hat{\gamma}$, as well as the pre-trial p-value.  
}
\label{tab:untriggerhighligths}

\resizebox{\textwidth}{!}{
\begin{tabular}{lrr|rrrrr|rrrrr}
\hline
\multicolumn{3}{c|}{Direction} & \multicolumn{10}{c}{Results} \\
\hline
& $\delta$ & $\alpha$ &
\multicolumn{5}{c|}{Gaussian-shaped time profile} &
\multicolumn{5}{c}{box-shaped time profile} \\
&  &  &
$\hat{T}_0$ & $\hat{\sigma}_t$ & $\hat{\mu}_\mathrm{sig}$ & $\hat{\gamma}$ & p-val &
$\hat{T}_0$ & $\hat{\sigma}_t$ & $\hat{\mu}_\mathrm{sig}$ & $\hat{\gamma}$ & p-val \\
& [$^\circ$] & [$^\circ$] & [MJD] & [days] & & & & [MJD] & [days] & & & \\
\hline
Hotspot (G) & 9.8 & 141.3 & 57666.5 & 1.9 & 3.0 & 2.4 & $8.3\times10^{-6}$ &  &  &  &  &  \\
Hotspot (B) & 9.4 & 141.1 &  &  &  &  &  & 57666.9 & 2.3 & 3.2 & 2.4 & $8.8\times10^{-6}$ \\
\hline
J0242+1101 & 11.02 & 40.64 & 56681.2 & 312.7 & 5.3 & 2.5 & $2.1\times10^{-2}$ & 56602.6 & 434.5 & 5.6 & 2.3 & $8.2\times10^{-3}$ \\
\hline
PKS~1502+106 & 10.49 & 226.10 & 55492.9 & 2.5 & 3.0 & 3.4 & $1.1\times10^{-2}$ & 55494.7 & 4.8 & 3.3 & 3.4 & $1.5\times10^{-2}$ \\
TXS~0506+056 & 5.69 & 77.35 & 56818.3 & 141.1 & 3.3 & 3.2 & $3.7\times10^{-2}$ & 56828.5 & 189.8 & 3.4 & 3.2 & $2.2\times10^{-2}$ \\
\hline
\end{tabular}
}
\end{table*}

\section{Conclusions} \label{sec:concl}

This work presents the results of an extensive search for astrophysical neutrino sources using the complete dataset collected by the ANTARES neutrino telescope over its 15 years of operation. The dataset, comprising both track-like and shower-like events, was analysed to identify both steady and flaring neutrino emission.

A comprehensive, time-integrated scan of the celestial sphere was performed to search for clustering of high-energy neutrino events. The most significant excess is located at $(\alpha, \delta) = (200.5^\circ\!, 17.7^\circ)$ and its significance corresponds to a post-trial p-value of 0.38, indicating no significant deviation from the background hypothesis.

A dedicated search for extended emission from the Galactic Plane ($|b| < 5^\circ$) was also carried out, motivated by recent IceCube observations of a Galactic neutrino flux, as well as by the HAWC and LHAASO detection of several extended Galactic \textit{PeVatrons}. No significant signal was found in this region.

In addition, an analysis of a list of known candidate neutrino sources was performed, including 161 point-like and eight extended objects.  The blazar MG3J225517+2409 emerged as the most significant source in this list, 
although still consistent with the background-only hypothesis. A local excess of $2.4\sigma$ was also observed from the direction of TXS0506+056, while no significant excess was detected from NGC~1068. The derived upper limits for this source remain compatible with the flux reported by IceCube.

The dataset was also examined for time-dependent flaring neutrino emissions across the entire sky. The most significant flare-like structure was found at $(\alpha, \delta) = (141.3^\circ\!, 9.8^\circ)$, with a Gaussian-shaped time profile centred at MJD~$57666.5$, a duration of $\sim$$4$ days, and a post-trial p-value of 0.30. A consistent result is found assuming a box-like profile.

 Lastly, directions previously highlighted by the IceCube Collaboration in a time-dependent flare search were investigated. Among the four considered sources—PKS1502+106, TXS0506+056, M87, and NGC1068—coincident flare activity is observed in the ANTARES data for PKS1502+106 and TXS~0506+056. The estimated chance probability of observing such overlaps by background fluctuations alone is of the order of 0.02\%, making the result statistically rare and therefore of particular interest.

\clearpage
\appendix

\section{Appendix} \label{Appendix}

{ 
\scriptsize
\setlength{\tabcolsep}{.3em}

\begin{longtable}{lccccccccccc}
    \caption[Results from the candidate list search]{\small List of analysed astrophysical point-like and extended objects. Reported are the source name, declination $\delta$, right ascension $\alpha$, source extension $\sigma_\text{Ext}$, best-fit number of signal events $\hat{\mu}_\mathrm{sig}$, pre-trial p-value and 90\% CL upper limits on the flux normalisation factor $\Phi_{\rm 1\,GeV}^{90\% \mathrm {CL}}$ for a $E^{-\gamma}$ spectrum with $\gamma = 2.0\,(2.5)$, in units of $10^{-8}\,(10^{-6})\,\times \rm{GeV^{-1} \, cm^{-2} s^{-1}}$. Sources for which a pre-trial significance over $2.0\sigma$ has been obtained are highlighted in bold. Dashes (--) in the fitted number of source events and pre-trial p-value indicate sources with null fitted signal. Dashes on the source extension indicate that the source is treated as point-like.} 
    
    \label{tab:LimitsFixAll} \\

   \multirow{3}{*}{Name} & 
   \multirow{3}{*}{$\delta [^{\circ}]$} &
   \multirow{3}{*}{$\alpha [^{\circ}]$} &
   \multirow{3}{*}{$\sigma_\text{Ext} \, [^{\circ}]$} & 
   \multicolumn{3}{c}{$\gamma = 2.0$} & ~ &
   \multicolumn{3}{c}{$\gamma = 2.5$} \\ \cline{5-7} \cline{9-11} \\[-6pt] 
    &  &  &   & 
    $\hat{\mu}_\mathrm{sig}$ & p-val & $\Phi_{\rm 1\,GeV}^{90\%} $ & ~ & 
    $\hat{\mu}_\mathrm{sig}$ & p-val & $\Phi_{\rm 1\,GeV}^{90\%} $ \\[2pt] \hline \\[-6pt]
    \endfirsthead
        \caption[]{\small [Continued] } \\

        \multirow{3}{*}{Name} & 
        \multirow{3}{*}{$\delta [^{\circ}]$} & 
        \multirow{3}{*}{$\alpha [^{\circ}]$} & 
        \multirow{3}{*}{$\sigma_\text{Ext} \, [^{\circ}]$} & 
        \multicolumn{3}{c}{$\gamma = 2.0$}  & ~ &
        \multicolumn{3}{c}{$\gamma = 2.5$}  \\ \cline{5-7} \cline{9-11} \\[-6pt] 
        &   &   &   & 
        $\hat{\mu}_\mathrm{sig}$ & p-val & $\Phi_{\rm 1\,GeV}^{90\%}$ & ~ &
        $\hat{\mu}_\mathrm{sig}$ & p-val & $\Phi_{\rm 1\,GeV}^{90\%}$ \\[2pt] \hline \\[-6pt]
    \endhead

    SMC & $-72.83$ & 13.18 & -- & -- & -- & 0.37 & ~ & -- & -- & 0.65 \\
    LMC N132D & $-69.50$ & 81.26 & -- & -- & -- & 0.37 & ~ & -- & -- & 0.85 \\
    Circinus & $-65.34$ & 213.29 & -- & --  &  --  & 0.58 & ~ & 0.4 & 0.25 & 1.01 \\
    PSR B1259-63 & $-63.83$ & 195.70 & -- & --  &  --  & 0.38 & ~ & -- & -- & 0.72 \\
    Danks 1 & $-62.69$ & 198.12 & 0.66 & -- & -- & 0.43 & ~ & -- & -- & 0.78 \\
    RCW86 & $-62.48$ & 220.68 & -- & -- & -- & 0.38 & ~ & -- & -- & 0.72 \\
    HESS J1507-622 & $-62.34$ & 226.75 & -- & -- & -- & 0.50 & ~ & -- & -- & 0.90\\
    NGC 3603 & $-61.25$ & 168.79 & -- & -- & -- & 0.47 & ~ & -- & --  & 0.77 \\
    ESO 139-G12 & $-59.94$ & 264.41 & -- & -- & -- & 0.38 & ~ & -- & -- & 0.72 \\
    SNR G318.2+00.1 & $-59.46$ & 224.42 & -- & -- & -- & 0.38 & ~ & -- & -- & 0.72 \\
    HESS J1503-582 & $-58.74$ & 226.46 & -- & -- & -- & 0.37 & ~ & -- & -- & 0.72 \\
    Westerlund 2 & $-57.78$ & 155.81 & -- & -- & -- & 0.46 & ~ & 0.3 & 0.26 & 1.10 \\ 
    CirX-1 & $-57.17$ & 230.17 &  --  &  -- & -- & 0.44 & ~ & -- & -- & 0.92 \\
    IC$_{hot spot}$ South & $-56.5$ & 350.2 & -- & 0.3 & 0.16 & 0.64 & ~ & 0.1 & 0.27 & 1.07 \\ 
    HESS J1614-518 & $-51.87$ & 243.54 &  --  &  -- & -- & 0.37 & ~ & -- & -- & 0.77 \\
    PKS 2005-489 & $-48.82$ & 302.37 & -- & 0.6 & 0.16 & 0.55 & ~ & 1.5 & 0.11 & 1.53 \\ 
    GX 339-4 & $-48.79$ & 255.70 &  -- & -- &  --  & 0.49  & ~ & 0.8 & 0.20 & 1.30 \\
    HESS J1641-463 & $-46.31$ & 250.26 & -- & 1.3 & 0.11 & 0.68 & ~ & 1.7 & 0.10 & 1.59 \\
    RX J0852.0-4622 & $-46.37$ & 133.00 & 0.63 & -- & -- & 0.45 & ~ & -- & -- & 0.89 \\
    Westerlund 1 & $-45.85$ & 251.76 & 0.80 & 1.5 & 0.11 & 0.79 & ~ & 2.0 & 0.14 & 1.84 \\ 
    \textbf{Vela X} & \textbf{$-45.60$} & \textbf{128.75} & \textbf{0.58} & \textbf{2.8} & \textbf{0.027} & \textbf{0.97} & ~ & 3.2 & 0.040 & 2.24 \\
    PKS 0537-441 & $-44.08$ & 84.71 & -- & -- & -- & 0.42 & ~ & -- & -- & 0.82 \\
    Centaurus A & $-43.02$ & 201.36 & -- & 1.0 & 0.064 & 0.78 & ~ & 1.2 & 0.075 & 1.84 \\
    PKS 1424-418 & $-42.11$ & 216.99 & -- & -- & -- & 0.73 & ~ & -- & -- & 0.82\\
    1ES 2322-409 & $-40.66$ & 351.20 & -- & --  &  --  & 0.39 & ~ & -- & -- & 0.83 \\
    J0106-4034 & $-40.57$ & 16.69 & -- & -- & -- & 0.47 & ~ & 0.5 & 0.058 & 1.25 \\
    RX J1713.7-3946 & $-39.75$ & 258.25 &  --  &  -- & -- & 0.42 & ~ & --& --& 0.98 \\
    CTB 37A & $-38.52$ & 258.56 & -- & -- & -- &  0.51 & ~ & 0.3 & 0.042 & 1.23 \\   
    PKS 0426-380 & $-37.93$ & 67.17 & -- & --  &  --  & 0.39 & ~ & -- & -- & 0.82 \\
    PKS 1454-354 & $-35.67$ & 224.36 & -- & 0.3 & 0.17 & 0.73 & ~ & 1.3 & 0.14 & 1.72 \\
    PKS 0625-35 & $-35.49$ & 96.78 & -- & --  &  --  & 0.45  & ~ & -- & -- & 0.91 \\ 
    TXS 1714-336 & $-33.70$ & 259.40 & -- & --  &  --  & 0.56 & ~ & -- & -- & 1.15 \\  
    Swift J1656.3-3302 & $-33.04$ & 254.07 & -- & --  &  --  & 0.45 & ~ & -- & -- & 0.91\\
    PKS 0548-322 & $-32.27$ & 87.67 & -- & --  &  --  & 0.45 & ~ & -- & -- & 0.98 \\
    HESS J1746-308 & $-30.84$  & 266.57  & -- & -- & -- & 0.83 & ~ & -- & -- & 0.91 \\
    H2356-309 & $-30.63$ & 359.78 & -- & --  &  --  & 0.45  & ~ & -- & -- & 0.91 \\
    HESS J1741-302 & $-30.38$ & 265.32 & -- & 0.8 & 0.12 & 0.80 & ~ & 1.4 & 0.095 & 1.91 \\
    PKS 2155-304 & $-30.22$ & 329.72 & -- & --  &  --  & 0.45  & ~ & -- & -- & 0.91  \\  
    PKS 1622-297 & $-29.90$ & 246.50 & -- & --  &  --  & 0.48  & ~ & -- & -- & 0.95  \\ 
    J1924-2914 & $-29.24$ & 291.21 & -- & -- & -- & 0.69  & ~ & 0.2 & 0.26 & 1.42  \\    
    \textbf{Galactic Centre} & \textbf{$-29.01$} & \textbf{266.43} & -- &\textbf{2.1} & \textbf{0.017} & \textbf{1.18} & ~ & \textbf{2.4} & \textbf{0.012} & \textbf{2.82} \\ 
    J2258-2758 & $-27.97$ & 344.52 & -- & -- & -- & 0.48 & ~ & -- & -- & 0.95 \\
    J1625-2527 & $-25.46$ & 246.45 & -- & -- & -- & 0.48 & ~ & -- & -- & 0.95 \\
    V46741 Sgr & $-25.40$ & 274.86 & -- & -- & -- & 0.48 & ~ & -- & -- & 0.95 \\
    NGC 253 & $-25.29$ & 11.88 & -- & -- & -- & 0.56  & ~ & -- & -- & 1.11 \\    
    Terzan5 & $-24.90$ & 266.95 &  --  &  -- & -- & 0.70  & ~ & 1.0 & 0.17 & 1.68 \\
    1ES 1101-232 & $-23.49$ & 165.91 & -- & -- & -- & 0.50 & ~ & -- & -- & 0.98 \\ 
    J0457-2324 & $-23.24$ & 270.43 & -- & -- & -- & 0.51  & ~ & -- & -- & 1.15  \\ 
    W28 & $-23.34$ & 270.43 & -- & --  &  --  & 0.65 & ~ & 0.2 & 0.26 & 1.49 \\   
    J1833-210A & $-21.06$ & 278.42 & -- & -- & -- & 0.51 & ~ & -- & -- & 0.98 \\
    J0836-2016 & $-20.28$ & 129.16 & -- & -- & -- & 0.71 & ~ & -- & -- & 1.2 \\
    J1911-2006 & $-20.12$ & 287.79 & -- & -- & -- & 0.50 & ~ & -- & -- & 0.98 \\
    eHWC J1809-193 & $-19.34$ & 272.46 &  --  &  -- & -- & 0.51 & ~ & -- & -- & 0.98 \\
    \textbf{J0609-1542} & \textbf{$-15.71$} & \textbf{92.42} & -- & \textbf{1.2}  & \textbf{0.0073} & \textbf{1.41} & ~ & \textbf{1.3} & \textbf{0.018} & \textbf{2.93} \\  
    SNR G015.4+00.1 & $-15.47$ & 274.52 & -- & -- & -- & 0.59  & ~ & 0.5 & 0.23 & 1.65 \\
    J2158-1501 & $-15.02$ & 329.53 & -- & -- & -- & 0.52 & ~ & -- & -- & 1.04\\      
    LS 5039 & $-14.83$ & 276.56 & -- & -- & -- & 0.52 & ~ & -- & -- & 1.04\\   
    LHAASO J1825-1337u & $-13.63$ & 276.45 & -- & -- & -- & 0.63 & ~ & -- & -- & 1.38 \\   
    QSO 1730-130 & $-13.10$ & 263.30 & -- & 1.3 & 0.077 & 1.00 & ~ & 1.4 & 0.10 &2.14 \\    
    J1337-1257 & $-12.96$ & 204.42 & -- & -- & -- & 0.52 & ~ & -- & --& 1.04\\    
    J2246-1206 & $-12.11$ & 241.58 & -- & -- & -- & 0.52 & ~ & -- & -- & 1.04\\   
    1ES 0347-121 & $-11.99$ & 57.35 & -- & -- & -- & 0.52 & ~ & -- & -- & 1.04\\    
    PKS 0727-11 & $-11.70$ & 112.58 & -- & 1.4 & 0.041 & 1.13 & ~ & 1.5 & 0.072 & 2.31 \\
    TXS 1749-101 & $-10.18$ & 268.15 & -- & -- & -- & 0.54 & ~ & -- & -- & 1.11 \\
    HESS J1828-099 & $-9.99$ & 277.24 & -- & 0.8 & 0.076 & 1.05 & ~ & 1.2 & 0.047 & 2.69  \\   
    J1512-0905 & $-9.10$ & 228.21 & -- & -- & -- & 0.54 & ~ & -- & -- & 1.11 \\          
    HESSJ 1834-087 & $-8.76$ & 278.69 &  --  &  -- & -- & 0.54 & ~ & -- & -- & 1.11 \\  
    J0607-0834 & $-8.58$ & -92.00 & -- &0.6 & 0.15 & 0.97 & ~ & 1.0 & 0.057 & 2.59 \\   
    PKS 1406-076 & $-7.90$ & 212.20 & -- & --  &  --  & 0.55 & ~ & -- & -- & 1.37 \\
    QSO 2022-077 & $-7.60$ & 306.40 & -- & 0.4 & 0.13 & 0.92 & ~ & 0.6 & 0.17 & 1.94 \\      
    RS Ophiuchi & $-6.71$ & 267.55 & -- & 1.1 & 0.031 & 1.22 & ~ & 1.4 & 0.043 & 2.73 \\    
    J0006-0623 & $-6.39$ & 1.56 & -- & 1.5 & 0.033 & 1.20 & ~ & 1.5 & 0.049 & 2.67 \\     
    3C279 & $-5.79$ & 194.05 & -- & 1.4 & 0.065 & 1.07 & ~ & 1.4 & 0.090 & 2.35 \\       
    1LHAASO J1839-0548u & $-5.79$ & 279.79 & -- &  --  &  --  & 0.54 & ~ & -- & -- & 1.26 \\  
    J2225-0457 & $-4.95$ & 336.45 & -- & -- & -- & 0.55 & ~ & -- & -- & 1.17 \\
    4FGL J0307.8-0419 & -$4.33$ & 46.95 & -- & -- & -- & 0.55 & ~ & -- & -- & 1.42\\
    \textbf{PKS 1741-038} & \textbf{$-3.83$} & \textbf{266.00} & -- & 1.6 & 0.038 & 1.20 & ~ & \textbf{3.3} & \textbf{0.012} & \textbf{3.52} \\       
    1LHAASO J1843-0335u & $-3.60$ & 280.91 & -- & --  &  --  & 0.55 & ~ & -- & -- & 1.17 \\
    HESS J1848-018 & $-1.89$ & 282.12 & -- & -- & -- & 0.56 & ~ & -- & -- & 1.18 \\   
    J0339-0416 & $-1.78$ & 54.88 & -- & -- & -- & 0.56 & ~ & -- & -- & 1.18 \\    
    J0423-0120 & $-1.34$ & 65.82 & -- & -- & -- & 0.70 & ~ & 0.5 & 0.22 & 1.88 \\   
    J0725-0054 & $-0.92$ & 111.46 & -- & -- & -- & 0.56 & ~ & -- & -- & 1.18 \\   
    1LHAASO J1848-0001u & $-0.02$ & 282.19 & -- & -- & -- & 0.56 & ~ & -- & -- & 1.18 \\    
    NGC 1068 & $-0.01$ & 40.67 &  --  &  --  & -- & 0.7 & ~ & 0.1 & 0.26 & 1.76 \\ 
    J2136+0041 & 0.70 & 324.16 & -- & 0.2 & 0.18 & 0.91 & ~ & 1.3 & 0.10 & 2.60 \\  
    3HWC J1852+013 & 1.34 & 283.05 & -- & -- & -- & 0.57 & ~ & -- & -- & 1.30 \\
    J1058+0133 & 1.57 & 164.62 & -- & -- & -- & 0.66 & ~ & -- & -- & 1.45 \\   
    J0108+0135 & 1.58 & 17.16 & -- & -- & -- & 0.57 & ~ & -- & -- & 1.30 \\   
    PKS 0736+017 & 1.79 & 28.17 & -- & -- & -- & 0.57 & ~ & -- & -- & 1.30 \\ 
    PKS 0215+015 & 1.75 & 34.45 & -- & -- & -- & 0.97 & ~ & 1.8 & 0.10 & 2.61 \\
    RGB J0152+017 & 1.79 & 28.17 &  --  &  -- & -- & 0.57 & ~ & -- & -- & 1.26 \\
    J1229+0203 & 2.05 & 187.50 & -- & -- & -- & 0.60 & ~ & -- & -- & 1.27\\
    HESS J1858+020 & 2.06 & 284.57 & -- & -- & -- & 0.57 & ~ & -- & -- & 1.27 \\
    TXS 0310+022 & 2.50 & 48.30 & -- & -- & -- & 0.57 & ~ & -- & --  & 1.26 \\
    \textbf{3C403} & \textbf{2.51} & \textbf{298.07} & -- & \textbf{2.5} & \textbf{0.00048} & \textbf{2.02} & ~ & \textbf{2.6} & \textbf{0.00047} & \textbf{5.34} \\
    CGCG 420-015 & 4.03 & 73.38 & -- & -- & -- & 0.58 & ~ & -- & --& 1.67 \\
    SS433 & 4.98 & 287.96 & -- & -- & -- & 0.58 & ~ & --& --& 1.27\\
    J0433+0521 & 5.35 & 68.30 & -- & -- & -- & 0.58 & ~ & --& --& 1.27 \\
    \textbf{TXS 0506+056 }& \textbf{5.69} & \textbf{77.35} & -- & \textbf{2.2} & \textbf{0.0075} & \textbf{1.6} & ~ & \textbf{2.6} & \textbf{0.0087} & \textbf{4.01} \\
    HESS J0632+057 & 5.81 & 98.24 & -- & 1.4 & 0.033 & 1.31 & ~ & 2.2 & 0.032 & 3.50 \\
    1LHAASO J1908+0615u & 6.26 & 287.05 & -- & -- & -- & 0.58 & ~ & -- & -- & 1.61 \\
    PKS 2145+067 & 6.96 & 327.02 & -- & --  & -- & 0.59 & ~ & -- & --  & 1.34 \\  
    B1030-074 & 7.19 & 158.39 & -- & -- & -- & 5.9 & ~ & -- & -- & 1.34\\
    3HWC J1907+085 & 8.57 & 286.79 & -- & -- & -- & 0.59 & ~ & -- & -- & 1.34 \\
    W 49B & 9.09 & 287.78 & -- & 1.0 & 0.039 & 1.27 & ~ & 1.0 & 0.056 & 3.17\\
    OT 081 & 9.65 & 267.89 & -- & -- & -- & 0.90 & ~ & 0.4 & 0.21 & 2.21 \\
    HESS J1921+101 & 10.15 & 288.21 & -- & -- & -- & 0.6 & ~ & --&-- & 1.45\\
    PKS 1502+106 & 10.49 & 226.10 & -- & -- & -- & 0.58 & ~ & --&-- & 1.36\\
    \textbf{J0242+1101} & \textbf{11.02} & \textbf{40.64} & -- & \textbf{3.7} & \textbf{0.0074} & \textbf{1.58} & ~ & \textbf{5.3} & \textbf{0.0039} & \textbf{4.63} \\
    1LHAASO J1959+1129u & 11.49 & 299.82 & -- & -- & -- & 0.59 & ~ & -- & --& 1.34\\
    RBS 0723 & 11.56 & 131.80 & -- & -- & -- & 0.72 & ~ & --& --& 1.69\\
    3HWC J1914+118 & 11.72 & 288.68 & -- & -- & -- & 0.6  & ~ &-- &-- &1.44\\
    J2232+1143 & 11.73 & 338.15 & -- & -- &-- & 0.61 & ~ & --& --& 1.44\\
    J0121+1149 & 11.83 & 20.42 & -- & -- & -- & 0.61 & ~ & --& --& 1.44 \\
    J1230+1223 & 12.39 & 187.71 & -- & 0.4 & 0.17 & 1.01 & ~ & 0.6 & 0.17 & 2.54 \\
    J0750+1231 & 12.52 & 117.72 & -- & 1.6 & 0.084 & 1.14  & ~ & 1.9 & 0.084 & 3.11\\
    PKS 1413+135 & 13.35 & 214.03 & -- & -- & -- & 0.61  & ~ & --&-- & 1.44\\
    J0530+1331 & 13.53 & 82.74 & -- & -- & -- & 0.82 & ~ & --&-- & 1.73\\ 
    W 51 & 14.14 & 290.75 & -- & -- & -- & 0.61 &  ~ &--&-- & 1.44\\
    VER J0648+152 & 15.27 & 102.20 & -- & -- & -- & 0.71 & ~ & -- & -- & 1.76\\
    J2253+1608 & 16.15 & 343.49 & -- & -- & -- & 0.62 & ~ & -- & -- & 1.44\\
    PKS 0235+164 & 16.61 & 39.66 & -- & -- & -- & 0.62 & ~ & -- & -- & 1.44\\
    PKS 0735+178 & 17.71 & 114.53 & -- & -- & -- & 0.62 & ~ & -- & -- & 1.44\\
    1LHAASO J1928+1813u & 18.23 & 292.07 & 0.63 &  -- & -- & 1.01 & ~ & 0.27 & 0.28 & 2.42 \\
    J0854+2006 & 20.11 & 133.70 & -- & -- & -- & 0.63  & ~ & -- & -- & 1.91 \\
    RGB J2243+203 & 20.35 & 340.98 & -- & 0.7 & 0.11 & 1.15 & ~ & 0.8 & 0.12 & 3.07 \\
    VER J0521+211 & 21.21 & 80.40 & -- & -- & -- & 0.80 & ~ & 0.8 & 0.16 & 2.82 \\
    4C+21.35 & 21.38 & 186.23 & -- & -- & -- &  0.63 & ~ & -- & -- & 1.56\\
    Crab & 22.01 & 83.63 & -- & -- & -- & 0.70 & ~ & --&-- & 1.94\\ 
    IC 443 & 22.65 & 94.21 & -- & -- & -- & 0.75 & ~ & --&-- & 2.21\\
    S2 0109+22 & 22.74 & 18.02 & -- & -- & -- & 0.73 & ~ & --&-- & 1.84\\
    B1422+231 & 22.93 & 216.76 & -- & -- & -- & 0.81 & ~ & --&-- & 2.23\\
    ARP 220 & 23.51 & 233.74 & -- & -- & -- & 0.63 & ~ & --&-- & 1.78\\
    3HWC J1940+237 & 23.77 & 295.05 & -- & -- & -- & 0.66 & ~ & --&-- &1.78 \\
    PKS 1424+240 & 23.80 & 21.76 & -- & -- & -- & 0.73 & ~ & -- & -- & 1.78 \\
    \textbf{MG3 J225517+2409} & \textbf{24.19} & \textbf{343.82} & -- & \textbf{4.0} & \textbf{0.00024} & \textbf{2.28} & ~ & \textbf{4.8} & $\mathbf{6.4\cdot10^{-5}}$ & \textbf{8.20} \\
    3HWC J1950+242 & 24.26 & 297.69 & -- & 0.5 & 0.11 & 1.13 & ~ & 0.5 & 0.16 & 3.09 \\
    MS1221.8+2452 & 24.61 & 186.10 & -- & -- & -- & 0.66 & ~ & -- & -- & 1.78 \\
    PKS 1441+25 & 25.03 & 220.99 & -- & 0.6 & 0.12 & 1.12 & ~ & 0.9 & 0.12 & 3.33 \\
    1ES 0647+250 & 25.05 & 102.69 & -- & -- & -- & 0.88 & ~ & 0.4 & 0.17 & 3.03 \\
    S3 1227+25   & 25.30 & 187.56 & -- & -- & -- & 0.66 & ~ & -- & -- & 1.78 \\ 
    3HWC J1951+266 & 26.61 & 297.90 & 1.7 & -- & -- & 0.93 & ~ & -- & -- & 2.90 \\
    W Comae     & 28.23 & 185.38 & -- & -- & -- & 0.66 & ~ & -- & -- & 1.78 \\
    1LHAASO J1959+2846u & 28.78 & 299.78 & -- & -- & --& 0.43 & ~ & 0.2 & 0.21 & 2.78 \\
    J0237+2848 & 28.80 & 39.47 & -- & -- & -- & 0.83 & ~ & --&-- & 1.78 \\         
    TON0599 & 29.24 & 179.88 & -- & -- & -- &  0.66 & ~ & --&-- & 1.78 \\ 
    1ES 1215+303 & 30.10 & 184.45 & -- & -- & -- & 0.72 & ~ & --&-- & 2.08\\
    1ES 1218+304 & 30.19 & 185.36 & --  & -- & -- & 0.72 & ~ & --&-- & 2.10\\
    B21811+31 & 31.74  & 273.40 &  -- & -- & -- & 0.72 & ~ & --&-- & 2.10\\
    J1310+3220 & 32.25 & 197.62 & --  & -- & -- & 0.83 & ~ & --&-- & 2.19\\
    B21420+32 & 32.39 & 215.63 & -- & -- & -- & 0.88 & ~ & --&-- & 2.74\\   
    1LHAASO J2002+3244u & 32.74 & 300.64 & -- & -- & -- & 1.02 & ~ & 0.6 & 0.14 & 3.73\\
    1LHAASO J2028+3352 & 33.88 & 307.21 & 1.70 &  --  & -- & 0.52 & ~ & -- &-- & 1.85 \\ 
    3HWC J2006+340 & 34.00 & 301.73 & -- & --  & -- & 0.85 & ~ & 0.1 & 3.2 & 3.20\\
    J1613+3412 & 34.21 & 243.42 & -- & -- & -- & 0.86 & ~ & -- & -- & 2.38\\
    S3 0218+35 & 35.94 & 35.27 & -- & --  & -- & 0.72 & ~ & -- & -- & 2.08\\
    1LHAASO J2018+3643u & 36.72 &  304.65 & -- & 0.17  & 0.16 & 1.12 & ~ & 0.35 & 0.19 & 3.41 \\
    1LHAASO J2027+3657u & 36.95 & 306.88 & -- & --  & -- & 0.99 & ~ & 0.22 & 0.29 & 3.46 \\
    J2015+3710 & 37.18 & 303.87 & -- & --  & -- & 1.16 & ~ & 0.03 & 0.26 & 3.74 \\       
    Mkn 421 &  38.21 & 166.11 & --  & -- & -- & 0.80 & ~ & -- & --& 2.52 \\
    B3 2247+381 & 38.43 & 342.53 & --  & -- & -- & 0.80 & ~ & -- & -- & 2.52 \\ 
    J0927+3902 & 39.04 & 141.76 & -- & 0.1  & 0.16 & 1.22 & ~ & 0.3 & 0.21 & 4.03 \\
    NGC 4151 & 39.41 & 182.63 & -- & -- & -- & 0.80 & ~ & -- & -- & 2.52\\
    Mkn 501 & 39.76 &  253.47 & -- & --  & -- & 0.80 & ~ & -- & -- & 2.52\\
    J1642+3948 & 39.81 & 250.75 & -- & -- & -- & 0.80 & ~ & -- & -- & 2.52\\
    J0555+3948 & 39.81 & 88.88  &  -- & --  & -- & 0.82 & ~ & -- & -- & 3.60\\
    CygOB2 & 41.04 & 307.44 & 2.16 & -- & -- & 0.59 & ~ & -- & --  & 2.43 \\ \hline

\end{longtable}

}

\newpage
\section*{Acknowledgements}
The authors acknowledge the financial support of the funding agencies:
Centre National de la Recherche Scientifique (CNRS), Commissariat \`a
l'\'ener\-gie atomique et aux \'energies alternatives (CEA),
Commission Europ\'eenne (FEDER fund and Marie Curie Program),
LabEx UnivEarthS (ANR-10-LABX-0023 and ANR-18-IDEX-0001),
R\'egion Alsace (contrat CPER), R\'egion Provence-Alpes-C\^ote d'Azur,
D\'e\-par\-tement du Var and Ville de La
Seyne-sur-Mer, France;
Bundesministerium f\"ur Bildung und Forschung
(BMBF), Germany; 
Istituto Nazionale di Fisica Nucleare (INFN), Italy;
Nederlandse organisatie voor Wetenschappelijk Onderzoek (NWO), the Netherlands;
Ministry of Education and Scientific Research, Romania;
MCIN for PID2021-124591NB-C41, -C42, -C43, PID2024-156285NB-C41, -C42, -C43 and PDC2023-145913-I00 funded by MCIN\-/\-AEI\-/\-10.13039/501100011033 and by “ERDF A way of making Europe”, for ASFAE/2022/014 and ASFAE/2022/023 with funding from the EU NextGenerationEU (PRTR-C17.I01) and Generalitat Valenciana, for Grant AST22\_6.2 with funding from Consejer\'{\i}a de Universidad, Investigaci\'on e Innovaci\'on and Gobierno de Espa\~na and European Union – NextGenerationEU, for CSIC-INFRA23013 and for CNS2023-144099, Generalitat Valenciana for CIDEGENT\-/2020/049, CIDEGENT/2021/23, CIDEIG/2023/20, ESGENT2024/24, CIPROM\-/2023/51, GRISOLIAP/2021/192 and INNVA1/2024/110 (IVACE+i), Spain;
Ministry of Higher Education, Scientific Research and Innovation, Morocco, and the Arab Fund for Economic and Social Development, Kuwait.
We also acknowledge the technical support of Ifremer, AIM and Foselev Marine
for the sea operation and the CC-IN2P3 for the computing facilities.

\bibliographystyle{utphys}
\bibliography{refs}{}

\end{document}